\documentclass[twocolumn]{aastex631}
\usepackage{graphicx}
\usepackage{epstopdf}
\usepackage{natbib}
\usepackage{graphicx}
\usepackage{natbib}
\bibpunct{(}{)}{;}{a}{}{,}
\usepackage{booktabs}
\usepackage{array}

\newcommand{\ee}[1]{\mbox{${} \times 10^{#1}$}}
\newcommand{\eten}[1]{\mbox{$10^{#1}$}}

\newcommand{\degree}{\mbox{$^{\circ}$}}

\newcommand{\kms}{\mbox{km s$^{-1}$}}
\newcommand\cmv{\mbox{cm$^{-3}$}}
\newcommand\cmc{\mbox{cm$^{-2}$}}

\newcommand{\x}{\mbox{${}\times{}$}}

\newcommand{\sfr}{\mbox{$\rm{SFR}$}}
\newcommand{\sfrmc}{\mbox{$\rm{SFR}_{\rm mc}$}}
\newcommand{\sfrseventy}{\mbox{${\rm SFR}_{\rm 70}$}}
\newcommand{\lseventy}{\mbox{$L_{\rm 70}$}}
\newcommand{\fseventy}{\mbox{$S_{\rm 70}$}}
\newcommand{\sigmasfr}{\mbox{$\Sigma_{\rm SFR}$}}
\newcommand{\rgal}{\mbox{$R_{\rm Gal}$}}

\newcommand{\mcat}{\mbox{$M_{\rm cat}$}}

\newcommand{\msun}{\mbox{M$_\odot$}}

\newcommand{\tk}{\mbox{$T_K$}}

\newcommand{\vlsr}{\mbox{$v_{\rm LSR}$}}
\newcommand{\n}{\mbox{$n$}}

\newcommand{\mmol}{\mbox{$M_{\rm mol}$}} 

\newcommand{\mean}[1]{\mbox{$\langle#1\rangle$}} 
\newcommand{\av}{\mbox{$A_V$}} 

\newcommand{\alphavir}{\mbox{$\alpha_{\rm vir}$}} 
\newcommand{\alphaco}{\mbox{$\alpha_{\rm CO}$}} 
\newcommand{\alphacounit}{\mbox{\msun (K\ \kms\ pc$^2$)$^{-1}$}}
\newcommand{\epsff}{\mbox{$\epsilon_{\rm ff}$}} 
\newcommand{\tff}{\mbox{$t_{\rm ff}$}} 
\newcommand{\tdep}{\mbox{$t_{\rm dep}$}} 
 
\newcommand{\hii}{\mbox{\ion{H}{2}}}

\newcommand{\jj}[2]{\mbox{$J = #1\rightarrow#2$}}

\newcommand{\go}{\mbox{$G_0$}}

\newcommand{\msunyr}{\mbox{M$_\odot$ yr$^{-1}$}}
\newcommand{\msunpc}{\mbox{M$_\odot$ pc$^{-2}$}}

\newcommand{\sfrtotpacs}{1.42}

\newcommand{\esfrtotpacsdexplus}{0.63}
\newcommand{\esfrtotpacsdexminus}{0.44}
\newcommand{\liconvtolicq}{0.915}
\newcommand{\lioverlicq}{0.86}
\newcommand{\averatio}{0.86}
\newcommand{\medratio}{0.64}
\newcommand{\nsourcesused}{29,996}
\newcommand{\methodratiomedian}{441}
\newcommand{\methodratiomedianhigh}{420}
\newcommand{\percentagehighsfr}{80}

\newcommand{\gtdexpone}{0.47}
\newcommand{\gtdexptwo}{0.051}
\newcommand{\egtdexpone}{0.01}
\newcommand{\egtdexptwo}{0.001}

\newcommand{\sfreliacorr}{1.46}
\newcommand{\sfrgianncorr}{1.76}
\newcommand{\esfreliacorr}{0.54}
\newcommand{\esfrgianncorr}{0.66}

\shorttitle {SFR in the Galaxy}
\shortauthors{Elia et al.}

\begin{document}

\title{Measuring Star Formation Rates in the Milky Way from Hi-GAL 70~\micron\ Observations}

\correspondingauthor{Davide Elia}
\email{davide.elia@inaf.it}

\author[0000-0002-9120-5890]{Davide Elia}
\affiliation{INAF-IAPS, Via del Fosso del Cavaliere 100, I-00133 Roma, Italy}

\author[0000-0001-5175-1777]{Neal J. Evans II}
\affiliation{Department of Astronomy, The University of Texas at Austin,
2515 Speedway, Stop C1400, Austin, Texas 78712-1205, USA}

\author[0000-0002-0294-4465]{Juan D. Soler}
\affiliation{INAF-IAPS, Via del Fosso del Cavaliere 100, I-00133 Roma, Italy}

\author[0000-0002-8757-9371]{Francesco Strafella}
\affiliation{Department of Mathematics and Physics `E. De Giorgi', University of Salento, Via per Arnesano, Lecce I-73100, Italy}
\affiliation{INFN, Sezione di Lecce, Via per Arnesano, Lecce I-73100, Italy}
\affiliation{INAF, Sezione di Lecce, Via per Arnesano, Lecce I-73100, Italy}

\author[0000-0003-1560-3958]{Eugenio Schisano}
\affiliation{INAF-IAPS, Via del Fosso del Cavaliere 100, I-00133 Roma, Italy}

\author[0000-0002-9826-7525]{Sergio Molinari}
\affiliation{INAF-IAPS, Via del Fosso del Cavaliere 100, I-00133 Roma, Italy}

\author[0000-0003-3869-6501]{Andrea Giannetti}
\affiliation{INAF - Istituto di Radioastronomia, Via P. Gobetti 101, I-40129 Bologna, Italy}

\author[0000-0002-3577-6488]{Sudeshna Patra}
\affiliation{Department of Physics, Indian Institute of Science Education and Research Tirupati, Yerpedu, Tirupati - 517619, Andhra Pradesh, India}

\begin{abstract}
Three methods for computing the total star formation rate of the Milky Way agree well with a reference value of $1.65\pm0.19$~\msunyr. They are then used to determine the radial dependence of the star formation rate and face-on map for the Milky Way. First, the method based on a model of star formation in Hi-GAL-defined dense clumps, adjusted for an increase in the gas-to-dust ratio with Galactocentric radius, predicts $1.65\pm0.61$~\msunyr. Second, the method using the 70~\micron\ emission, commonly used in other galaxies, with a technique to assign distances to the extended emission, predicts $\sfrtotpacs^{+\esfrtotpacsdexplus}_{-\esfrtotpacsdexminus}$~\msunyr. Finally, a method based on theoretical predictions of star formation efficiency as a function of virial parameter, with masses corrected for metallicity dependence, applied to a catalog of molecular clouds also predicts a value in agreement at $1.47$~\msunyr. The three methods predict the radial variation of the star formation rate, with remarkably good agreement from the CMZ out to about 20~kpc. More differences were seen in face-on maps with a resolution of 0.5~kpc made with the three approaches and in comparisons to the local (within 3~kpc) star formation rate, indicating limitations of the methods when applied to smaller scales. The 70~\micron\ star formation rate follows very closely the surface density of molecular gas, corrected for a metallicity-dependent CO conversion factor. A molecular gas depletion time of 1~Gyr is consistent with the data, as is a molecular Kennicutt-Schmidt relation with a power-law slope of $1.10\pm0.06$.
\end{abstract}



\section{Introduction}\label{intro}

One of the fundamental properties of a galaxy is its rate of star formation \citep{2014ARA&A..52..415M, 2020ARA&A..58..157T}. Consequently, several observational quantities have been used to estimate the star formation rate in other galaxies (e.g., \citealt{2012ARA&A..50..531K}). All of these quantities used for other galaxies are ultimately tied to indicators, like H$\alpha$, that trace only massive stars, making the results sensitive to assumptions about the initial mass function and timescales \citep{2024MNRAS.534.2426H}. Many more methods, some of which are less biased toward massive stars, can be used in the Milky Way \citep{2011AJ....142..197C}, and a remarkably precise value has been determined by \citet{2015ApJ...806...96L}. If certain technical problems can be overcome, the Milky Way can provide a calibration for methods used for other galaxies. In this paper, we focus on extending the use of the 70~\micron\ extragalactic estimator to the Milky Way. As a bonus, we add another method for calculating the star formation rate (SFR) of the Milky Way as a function of Galactocentric radius (\rgal), produce a top down view of the SFR surface density, and compare them to other calculations. These results provide a unified, global picture of star formation in the Milky Way, comparable to those available for other galaxies.

The Hi-GAL survey of the plane of the Milky Way \citep{2010PASP..122..314M} provided a data set that allows us to see the Milky Way in the context of other galaxies. The products include images in five photometric bands from 70 to 500~\micron\ \citep{2016A&A...591A.149M}, and the catalog of compact sources (typically corresponding to physical scales of clumps) with associated distances, masses, and luminosities \citep{2021MNRAS.504.2742E}. By ``clumps", we mean objects \citep[with typical sizes in the range 0.3-3~pc,][]{2007ARA&A..45..339B} generally associated with the formation of a group of stars, as opposed to a ``core," which would form a single stellar system
\citep{2007ARA&A..45..565M}. 

 \citet{2022ApJ...941..162E} applied to the Hi-GAL data set the method described by \citet{2013A&A...549A.130V,2017A&A...599A...7V} to compute SFRs based on clump masses (\sfrmc), according to the relationship,
\begin{equation}\label{sfrclump}
\left(\frac{\sfrmc}{\msunyr}\right) = 5.6\ee{-7} \left(\frac{\mcat}{\msun}\right)^{0.74},
\end{equation}
where \mcat\ is the catalog value for the clump mass (the element of the calculation). With this approach, depending on the selected sky area, one can estimate the SFR of single regions, or of the entire Milky Way. \citet{2022ApJ...941..162E} obtained a total star formation rate for the Milky Way of $(1.74 \pm 0.65)$~\msunyr\ by including every Hi-GAL star forming clump with an estimate of its heliocentric distance , increasing to $(1.96\pm 0.73)$~\msunyr\ if the contribution from sources without distance information is taken into account based on reasonable assumptions. 
These values are consistent with other recent estimates in the literature, as for example $(1.9\pm 0.4)$~\msunyr\ \citep{2011AJ....142..197C} or $(1.65\pm 0.19)$~\msunyr\
\citep{2015ApJ...806...96L}\footnote{In this work, previous SFR estimates were statistically combined through a hierarchical Bayesian approach to obtain this value. We use this result as a reference throughout this paper, and, for comparison with it, we express the obtained SFR estimates with two decimal digits as well.}.

The clump masses used by \citet{2022ApJ...941..162E} in Eq.~\ref{sfrclump} were obtained by \citet{2021MNRAS.504.2742E} by converting dust masses to total masses through a constant gas-to-dust ratio, $\gamma=100$, commonly assumed for the Solar neighborhood. Recent evidence suggests that $\gamma$ increases with Galactocentric radius, so we will reconsider the masses and SFR estimates from 
\citet{2022ApJ...941..162E}.

While Hi-GAL clump masses, and thus \sfrmc, are obtained from a multi-wavelength approach, namely a modified black body fit of their spectral energy distribution (SED), a monochromatic approach to the calculation of the SFR, at least for extragalactic sources, has been proposed too. 
\citet{2010ApJ...716..453L} found that the 70~$\micron$ band is the most suitable indicator of the SFR because it accounts for IR emission from both \hii\ regions and embedded star formation. 
\citet{2010ApJ...714.1256C} found a relationship between SFR and 70~\micron\ luminosity, based on 189 whole galaxies, calibrated on their previous SFR estimator using a combination of 24~$\micron$ and H$\alpha$ luminosities \citep{2007ApJ...666..870C}.
Their relation was applicable for oxygen abundances, $12 + \log[\rm{O/H}] > 8.1$, but became suspect for low (total galaxy) $\sfr < 0.1-0.3$~\msunyr\ or low 70~\micron\ luminosity, $\lseventy < 1.4\ee{42}$ erg s$^{-1}$.
Using 40 galaxies in the SINGS survey \citep{2003PASP..115..928K}, 
\citet{2010ApJ...725..677L} extended the relationship to regions {\it within} galaxies, 
finding good correlations down to 70~\micron\ fluxes, $\Sigma(70) \approx \eten{40}$ erg~s$^{-1}$~kpc$^{-2}$, surface densities of SFRs, 
$\Sigma_\mathrm{SFR} > 1\ee{-3}$~\msunyr\~kpc$^{-2}$ and oxygen abundances, $12 + \log[\rm{O/H}] > 8.4$. We use
\citet{2010ApJ...725..677L} for this investigation:
\begin{equation}\label{sfr70}
\left(\frac{\sfrseventy}{\msunyr}\right) = \left(\frac{\lseventy}{1.067\ee{43}~\rm{erg~s}^{-1}}\right)\,.
\end{equation}
We will explore the possible extrapolation of these relations to lower $\Sigma(70)$ and lower metallicity, using Hi-GAL data as estimators of the SFR, for comparison to the previous estimates and to provide perspective to extragalactic studies. 

The paper has a threefold aim:
\begin{itemize}
    \item To revise the \sfrmc\ obtainable from Eq.~\ref{sfrclump} by correcting upstream the input clump masses by taking into account a gas-to-dust ratio that varies with the Galactic location of the clump \citep[e.g.,][]{2017A&A...606L..12G}.
    \item To evaluate the applicability to the Milky Way of the prescription given by \citet{2010ApJ...725..677L} for external Galaxies, based on the emission at 70~\micron.
    \item To use the 70~\micron\ emission from Hi-GAL to compute the star formation rate and distribution over the Milky Way and compare them to the results from other methods.
\end{itemize}

In \S \ref{data}, we summarize the data used in our analysis. \S \ref{gtd} considers the effects of a gas-to-dust ratio that varies with Galactocentric radius on our previous estimates of SFRs. In \S \ref{internal}, we show that the star formation rate is extremely under-estimated when the 70~\micron\ emission of cataloged sources is used. \S \ref{compother} presents a solution by assigning distances to the extended emission. In \S\ref{disc} we compare the predictions from \S \ref{compother} for the total and distribution of SFRs to those of other methods. A summary of key results is provided in \S \ref{summary}.

\section{Data}\label{data}
We used data from Hi-GAL, an Open Time Key Project of the \textit{Herschel} satellite \citep{2010A&A...518L...1P} which delivered a complete and homogeneous mapping of the Galactic plane in five continuum far-infrared (FIR) bands: 70 and 160~\micron\ with the PACS camera \citep{2010A&A...518L...2P} and 250, 350, and 500~\micron\ with the SPIRE camera \citep{2010A&A...518L...3G}, respectively.

The data that play a central role in this article are the Hi-GAL 70~\micron\ maps \citep{2016A&A...591A.149M}. We also use the photometry of Hi-GAL compact sources at 70~\micron\ obtained by \citet{2016A&A...591A.149M} and \citet{2021MNRAS.504.2742E}. 

To apply the method of calculating the SFR from the clump masses, we also use the physical properties quoted by \citet{2021MNRAS.504.2742E}. Starting from the Hi-GAL compact source photometry in five bands, they built band-merged SEDs, and selected a a sub-sample of $\sim1.5\times 10^5$ of them, having a shape regular enough to undergo modified black body fitting \citep[e.g.,][]{eli16}, essentially a concave-down shape with fluxes available in at least three consecutive bands at $\lambda \geq 160$~\micron \citep[see][for details]{2017MNRAS.471..100E}. The fit procedure was performed on the portion of SEDs at $\lambda \geq 160$~\micron, since fluxes at 70~\micron\ are found to generally depart from a modified black body behavior. The presence of a 70~\micron\ detection is used, instead, to establish a first classification between star forming (70-\micron\ bright) and quiescent (70-\micron\ dark) clumps, respectively. Heliocentric distances were provided by \citet{2021A&A...646A..74M} for $\sim 80\%$ of the clumps. In particular, for the SFR calculation, only star forming clumps provided with a distance (around 30000 objects) are considered directly. 
\citet{2022ApJ...941..162E} showed a way to include clumps without distance (by assigning them random distances following the distribution of the available ones), that we applied as described in \S\ref{gtd}.

\section{Effects of adopting a non-constant gas-to-dust ratio}\label{gtd}

The derivation of the SFR through clump masses (Eq.~\ref{sfrclump}) requires a few assumptions to estimate the masses, as discussed by \citet{2022ApJ...941..162E}. Among those, the assumption of a gas-to-dust ratio  that is constant ($\gamma=100$) across the Milky Way, rather than a function of the Galactocentric radius, certainly affects the global Galactic SFR, and even more its radial profile and its 2D distribution in the Galactic plane. Therefore, before evaluating the utility of the 70~\micron\ emission as a SFR indicator, we reconsider the SFR derived by
\citet{2022ApJ...941..162E}
from the clump masses through Eq.~\ref{sfrclump} in light of evidence that the gas-to-dust ratio $\gamma$ varies with \rgal\ \citep{2017A&A...606L..12G}. Similar conclusions have been reached in studies of other galaxies \citep{2020ApJ...889..150A, 2020ApJ...897..184A, 2024arXiv240414503L}.

The detailed considerations are contained in Appendix \ref{appendixd2g}. We conclude that the strong dependence on \rgal\ found by \citet{2017A&A...606L..12G} would imply that the median surface density of clumps increases with \rgal. Because that result seems very unlikely, we adopt the following function, which produces a constant mean surface density of clumps:
\begin{equation}
    \log(\gamma_\mathrm{H}) = (\gtdexptwo \pm \egtdexptwo)\left(\frac{\rgal -R_0}{\rm{kpc}}\right) +2 \,,
\label{gtdelia}
\end{equation}
where $\gamma_\mathrm{H}$ indicates the function based on the full Hi-GAL sample, normalized to 100 at the solar circle ($R_0$). This gradient ($0.051$) is very consistent with the gradient in the elemental abundance of oxygen, as documented in Appendix~\ref{appendixd2g}.

In the rest of this paper, when comparing methods for estimating the SFR, for the clump mass-based method we will adopt the correction given by Eq.~\ref{gtdelia}. We use this relation to re-compute the total SFR of the Galaxy from the clump mass method, $\sfrmc = (1.46 \pm 0.54)$~\msunyr, and its radial dependence.
As done by \citet{2022ApJ...941..162E}, a 2D view of the SFR density in the Galaxy can be obtained too from the clump mass method. By applying the gas-to-dust ratio in Eq.~\ref{gtdelia}, we obtain maps comparable to the uncorrected map \citep[Figure~5 of][]{2022ApJ...941..162E} with a cylindrically symmetric exponential pattern described in those equations. We will discuss the corresponding SFR density map in \S\ref{sfr2d}.

All these determinations are obtained by considering Hi-GAL star forming clumps with a known mass, thanks to an available heliocentric distance estimate. \citet{2022ApJ...941..162E} described an approach for estimating a realistic contribution also from sources without a distance ($\sim 16\%$ of the total) to the whole Milky Way SFR (not applicable, however, to 1D or 2D mapping of SFR density). It consists in running Monte-Carlo realizations of virtual distances following the distribution of available distances. Assigning such distances to distanceless clumps, the corresponding masses were derived, and hence their individual contributions to SFR through Eq.~\ref{sfrclump}, as well as the total of these contributions. The average of this total over a number of realizations provides a reasonable additional term for the total SFR. Here we repeated the same procedure, but estimating also the simulated \rgal\ from the combination of source longitude (known) and distance (simulated), which enables us to apply Eq.~\ref{gtdelia}. 
In 10000 realizations carried out, the computed additional SFR term ranges from to 0.17 to 0.23~\msunyr, with an average of 0.19~\msunyr and a standard deviation of 0.005~\msunyr. This average value can be assumed, finally, as the term required for accounting for the distanceless clumps as well. A typical uncertainty of 0.07~\msunyr \citep[estimated as in][and dominating over the intrinsic variability of the SFR]{2022ApJ...941..162E} is found to be associated to these additional terms, and we assign it as the error bar for the average value.

\section{A first attempt: estimating the SFR from 70~\micron\ compact source photometry}\label{internal}

The \citet{2010ApJ...725..677L} method computes the SFR from the luminosity at 70~\micron\ through Eq.~\ref{sfr70}. A first attempt was made to apply this formula to the fluxes at 70~\micron\ of Hi-GAL compact sources quoted in the catalog of \citet{2021MNRAS.504.2742E}, converted to luminosities through the distance (where available), according to 

\begin{equation}\label{fluxtoluminosity}
\lseventy = \nu \fseventy \times 4 \pi d^2,
\end{equation}
where \fseventy\ is the flux density (per Hz), $\nu$ is the frequency corresponding to $\lambda=70~\micron$, and $d$ is the heliocentric distance. 


 Adding up the SFRs of all the individual catalog sources using Eq.~\ref{fluxtoluminosity} produces a total Milky Way $\sfr = 2.4\ee{-2}$~\msunyr, nearly two orders of magnitude too low. In fact, \citet{2013A&A...549A.130V}, who focused on only two $\sim 2^\circ \times 2^\circ$ Hi-GAL fields around $\ell=30^\circ$ and $59^\circ$, already found that such an approach leads to an underestimation, with respect to the clump mass method.

To understand this discrepancy, we then checked for possible systematic mistakes in the calibration of Hi-GAL maps at 70~\micron, by comparing them with IRAS ones at 60 and 100~\micron.
These considerations are detailed in Appendix \ref{appendix1}. 
We concluded that there is no calibration problem. 

As also detailed in Appendix \ref{appendix1}, we then checked whether the underestimations depended on the star formation rate, since the 70~\micron\ tracer is known to fail for low emission levels. Instead, we found that the underestimation persisted over the full range of star formation rates, as measured by the clump mass model.

We concluded that the 70~\micron\ emission assigned to individual sources drastically underestimates the SFR when a standard formula used for other galaxies is applied. This result is similar to what was found for 24~\micron\ emission by 
\citet{2013ApJ...765..129V}.

\section{Estimating the SFR from the whole emission in 70~\micron\ maps}\label{compother}

Once it is established that considering only the emission from 70~\micron\ compact sources leads to a severe underestimation of the SFR, we tried to get closer to the spirit underlying the approach of \citet{2010ApJ...725..677L}, which is applied to the emission from large regions of external galaxies, or even whole galaxies in the case of \citet{2010ApJ...714.1256C}. In this respect, we applied such a method to entire Hi-GAL maps, namely by counting both compact and diffuse emission (hereafter the ``whole" emission) from all pixels in the 70~\micron\ maps. 

To do that, the Hi-GAL $2\fdg2 \times 2\fdg2$ tiles at 70~\micron\ were re-projected onto 72 mosaics, each one extending over $5\degr$ in longitude (and the full extent in latitude), to avoid counting twice the pixels in the overlap zone between the original tiles \citep[typically $\sim 20\arcmin$,][]{2016A&A...591A.149M}. The pixel size was kept at 3\farcs2. The pixel intensities of these mosaics were converted into fluxes by converting units from MJy~sr$^{-1}$ to Jy~pixel$^{-1}$.

\begin{figure}[ht!]
\includegraphics[width=0.5\textwidth]{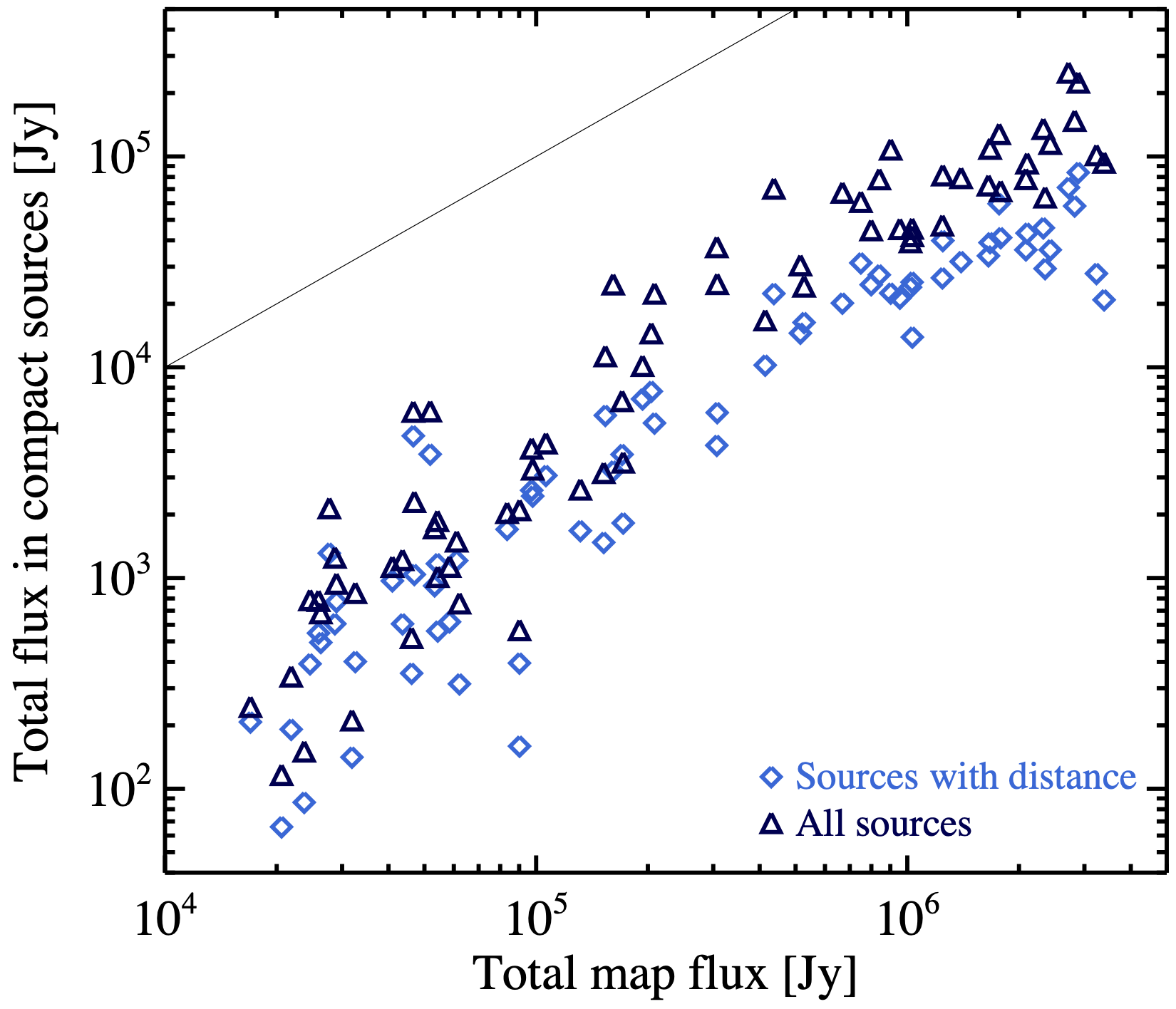}
\caption{For each of the 72 mosaics ($5^\circ$-wide in longitude) into which we have rearranged the Hi-GAL observations at 70~\micron\ of the whole Galactic plane, the total flux contained in the compact sources is plotted vs the flux given by the whole emission in the mosaic. In particular, light blue diamonds and dark blue triangles represent the total flux from 70~\micron\ sources provided with a heliocentric distance, and from all 70~\micron\ sources, respectively. The solid line represents the 1:1 relation.}
\label{tilesvscompact}
\end{figure}

In terms of observed fluxes, it can be seen (Figure~\ref{tilesvscompact}) that, in general, the amount of flux contained in compact sources present in a given mosaic represents a minority compared to the total emission from all map pixels, but it generally correlates with it. The statistics shown in Figure~\ref{tilesvscompact} are built with sources present in the ``filtered'' physical catalog of \citet{2021MNRAS.504.2742E} and provided with a heliocentric distance ($d$); for them, the fraction represented by the total flux in compact sources over that of the total map has a mean of $\sim 2\%$. Taking also into account the sources without distance information this fraction does not increase significantly. Even considering the totality of compact sources detected at 70~\micron\ by \citet{2016A&A...591A.149M}, which were partially filtered out by \citet{2021MNRAS.504.2742E}, the median ratio between their total flux and the whole emission in the mosaic is $\sim 4\%$. 

The fact that these strong differences in flux are similar to the gap observed between the SFR obtained from 70~\micron\ photometry of compact sources (\S\ref{internal}) and typical SFR estimates in the literature \citep[][and references therein]{2015ApJ...806...96L} further encourages us towards testing the approach based on considering the emission from all pixels of the 70~\micron\ maps. Of course, this presents the non-trivial problem of assigning a heliocentric distance to each pixel.

\begin{figure}[ht!]
\includegraphics[width=0.5\textwidth]{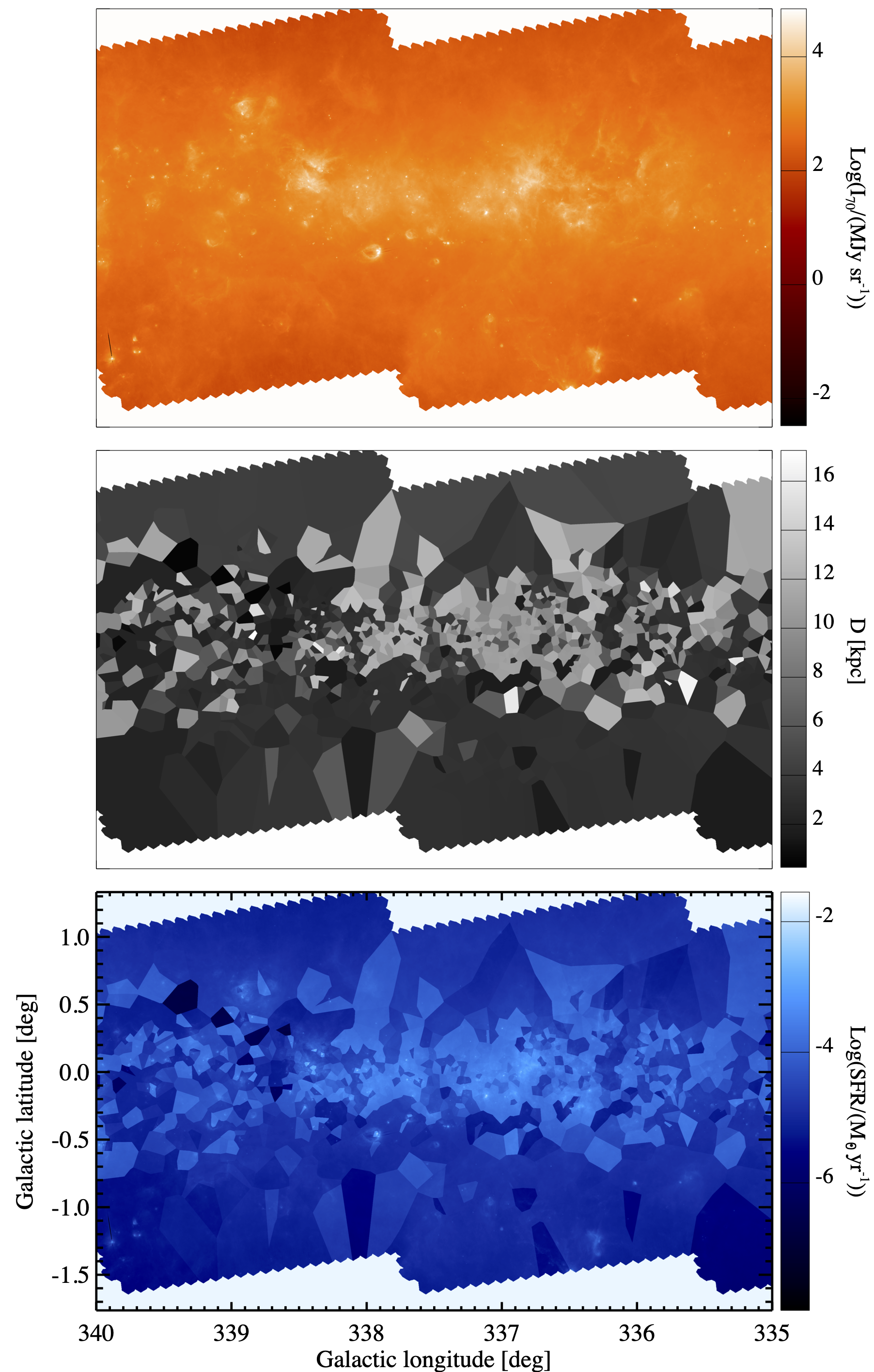}
\caption{Example of the application of the SFR estimation method based on considering the whole emission in Hi-GAL 70~\micron\ maps. \textit{Top}: 70~\micron\ mosaic in the $335^\circ-340^\circ$ longitude range, obtained by combining the original Hi-GAL tiles $\ell354$, $\ell352$, and $\ell349$. \textit{Middle}: map of the heliocentric distances assigned to pixels in the mosaic, based on the association to the closest (in the 2D map) Hi-GAL 70~\micron-bright source provided with a distance. \textit{Bottom}: SFR map obtained by combining the two previous panels, through Eq.~\ref{sfrpixel}. Galactic coordinates are shown in the bottom panel.}
\label{l70method}
\end{figure}

The procedure we adopted can be illustrated with the help of Figure~\ref{l70method}. 
The pixel intensities in the top panel were converted to fluxes, and then to luminosities through Eq.~\ref{fluxtoluminosity}, by assigning to each pixel the heliocentric distance of the closest (in terms of pixels) Hi-GAL 70~\micron\ source provided with a distance (Figure~\ref{l70method}, middle).
However, the presence of high-$|b|$ sources with very large distances can produce exceedingly large estimates of the total luminosity, since those portions of the maps are poorer in compact sources, so that very large areas are entirely assigned to such a large distance. The problem of these sources was addressed by \citet{2021A&A...646A..74M}, who took into account the criterion of \citet{1987ApJ...319..730S}, namely that a source cannot lie more than 140~pc above or below the midplane, as a possible method to solve the near/far distance ambiguity. However, they placed this criterion at the bottom of their decision tree, so that in many cases the far distance is definitely assigned before going all the way to invoking this criterion. On the contrary, we gave maximum priority to this requirement, bringing to the near distance estimate those sources not fulfilling it due to a far distance assignment (this applied to 1180 sources). 

The contribution of each pixel to the total SFR of the tile was evaluated through Eq.~\ref{sfr70} (Figure~\ref{l70method}, bottom). Summarizing, the conversion of the original intensity at 70~\micron\ ($I_{70}$) into the corresponding SFR per pixel ($\mathrm{SFR}_\mathrm{70pix}$) can be simply written, by combining all the conversion coefficients in Eqs.~\ref{sfr70} and~\ref{fluxtoluminosity}, as
\begin{equation}\label{sfrpixel}
\left(\frac{\mathrm{SFR}_\mathrm{70pix}}{\msunyr}\right) = 4.8\ee{-16}\, \left(\frac{I_{70}}{\mathrm{Jy~pixel}^{-1}}\right) \left(\frac{d}{\mathrm{pc}}\right)^2\:.
\end{equation}
We point out that, for accurate use of this equation, it may be necessary to first convert the units of the 70~\micron\ map to Jy/pixel.

All the 70~\micron\ emission in each tile was assigned the distance of the nearest catalog source and converted to a SFR for that source (\sfrseventy).

\begin{figure}[ht!]
\includegraphics[width=0.48\textwidth]{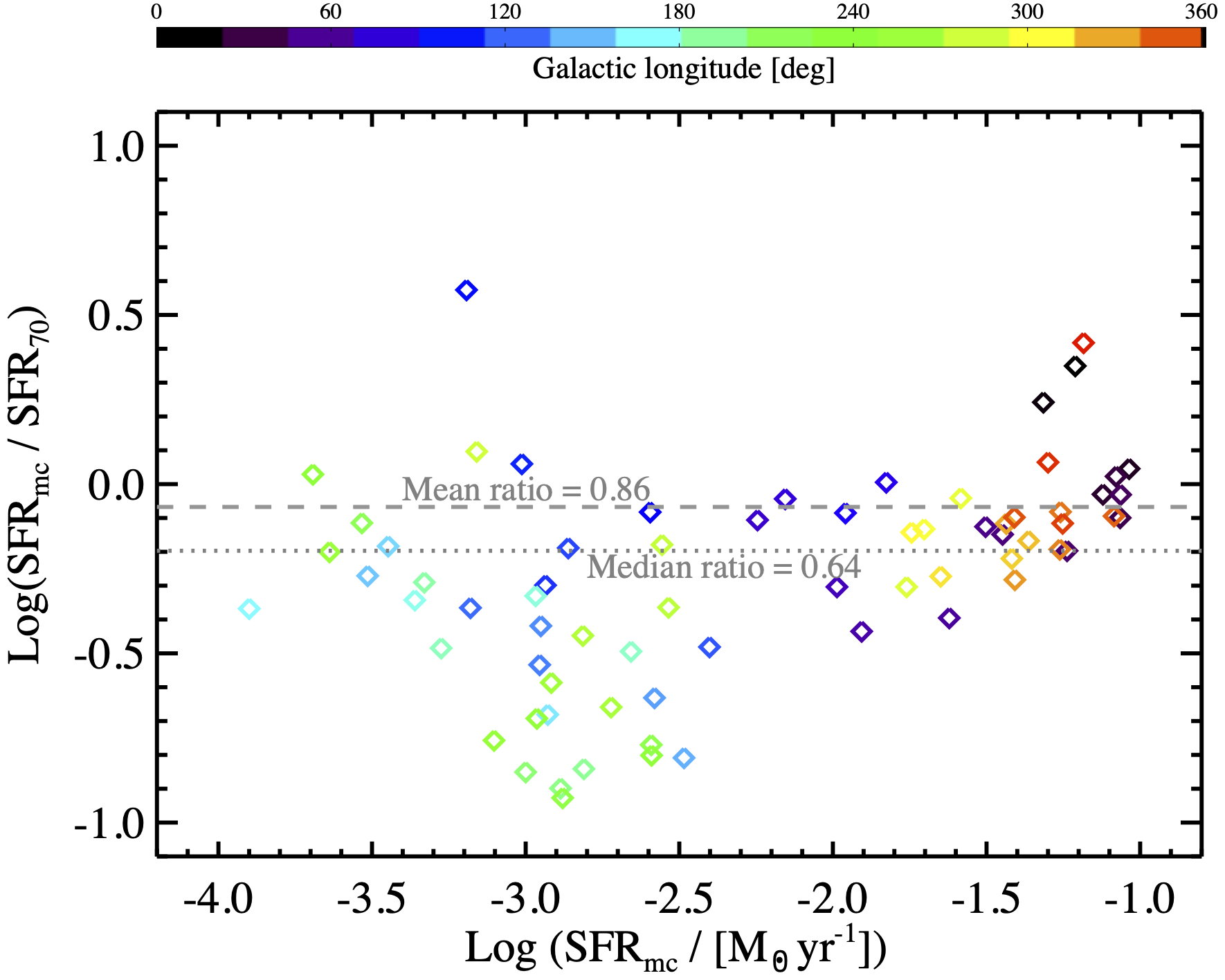}
\caption{
Comparison of SFR obtained for Hi-GAL mosaics (72 in total) through two different techniques. The ratio of the SFR$_{70}$ predicted by 70~\micron\, calculated by assigning the extended 70~\micron\ flux to the nearest source, to the SFR$_\mathrm{mc}$ from the clump mass model applied to all sources in the area is plotted versus the SFR$_\mathrm{mc}$. The symbol color is encoded by the characteristic longitude (the lower end of the longitude interval covered by each mosaic), according to the color bar shown on the top. The logarithms of the average and median ratio are represented as a grey dashed and dotted horizontal line, respectively.}
\label{tiles70vsmodel}
\end{figure}

This operation was repeated on the entire set of 72 re-projected Hi-GAL mosaics, together with the calculation of the SFR through clump counts (Eq.~\ref{sfrclump}) in the same areas. The comparison between the two estimates obtained for all the Hi-GAL mosaics is shown in Figure~\ref{tiles70vsmodel}. The whole emission at 70~\micron\ now accounted for, the two SFR estimates are now quite comparable: the median of $\sfrmc / \sfrseventy$ is \medratio, and the average is \averatio. As expected, the mosaics containing higher SFR are those located towards the inner Galaxy \citep[see also Figure~4 of][]{2022ApJ...941..162E}, and among these there are a few notable cases for which $\sfrmc$ is found to be remarkably larger than $\sfrseventy$. The opposite result is seen in many mosaics corresponding to the outer Galaxy, but at considerably lower values of SFR ($\ll 10^{-2}~\msunyr$).

\section{Comparison of SFR Methods}\label{disc}

From the results above, it is quite clear that using the luminosities in the Hi-GAL compact source catalog to compute SFRs will severely underestimate the star formation rate obtained from the clump mass model. The 70~\micron\ method will work only if the whole (extended plus compact) emission is used. Our method of assigning the 70~\micron\ emission a distance works well enough to get general agreement. We now proceed to compare the results of the two methods for the total, radial dependence, and 2D distributions of the star formation rate.

\subsection{Total SFR of the Galaxy}
Adding all the SFR based on 70~\micron\ emission for the entire Hi-GAL area now yields a total $\sfrseventy = (\sfrtotpacs^{+\esfrtotpacsdexplus}_{-\esfrtotpacsdexminus}$)~\msunyr. Following \citet{2010ApJ...725..677L}, the error bar estimation is not based on the uncertainty on the factor 1.067 in Eq.~\ref{sfr70} (which amounts to 0.017, i.e. 1.6\%), but rather on the dispersion of data in their diagram of 70~\micron\ brightness vs SFR, namely 0.16~dex (increasing to 0.18~dex when they consider low-metallicity galaxies as well), which corresponds to an asymmetric relative error bar of $+44\%$ and $-31\%$.

This total SFR is in remarkable agreement with that from the clump mass model: $\sfrmc = (\sfreliacorr \pm \esfreliacorr)$~\msunyr. However, this value of \sfrmc\ has been obtained using only clumps with a heliocentric distance estimate. As explained in Section~\ref{gtd}, an estimate of the possible contribution to the SFR by sources with unknown distance would further increase the total SFR to $(1.65 \pm 0.61)$~\msunyr.
In contrast, the SFR based on 70~\micron\ emission is calculated using all pixels of the survey. In this case, adding distance information for more clumps would not translate necessarily to additional star formation (as in the clump mass method), but rather in a re-assignment of distances for groups of pixels, which in turn can produce a decrease (/increase) of the SFR if the newly assigned distance is closer (/farther) than the one previously assumed. Therefore it is more appropriate to compare $\sfrseventy = (\sfrtotpacs^{+\esfrtotpacsdexplus}_{-\esfrtotpacsdexminus}$)~\msunyr\ with $\sfrmc = (1.69 \pm 0.62)$~\msunyr, which still agrees within the error bars.

The value of \sfrtotpacs~\msunyr is also very close to the total Galactic star formation rate of 1.65~\msunyr\ 
by \citet{2015ApJ...806...96L}, which encloses and summarises all the main results achieved up to that point. In this respect, to calibrate our result on this estimate, we suggest to correct to $\liconvtolicq \ee{43}~\rm{erg~s}^{-1}$ the factor in the equation in \citet{2010ApJ...725..677L} (Eq. \ref{sfr70} in this paper). This change is well within the uncertainties.

\begin{table*}
\begin{center}
\caption{Comparison among Milky Way star formation rates derived in this paper and values from the literature.\label{sfrsummary}}
\begin{tabular}{lccc}
\hline
 Method & SFR & SFR & Ref. \\
    & (Sources w/ distance) & (Sources w/+w/o distance) &  \\
        & [\msunyr] & [\msunyr] &  \\
\hline
\hline
Clump masses & $1.74 \pm 0.65$ & $1.96 \pm 0.73$ & \citet{2022ApJ...941..162E} \\
Clump masses corrected by Eq.~\ref{gianneq} & $\sfrgianncorr \pm \esfrgianncorr$ & $2.06 \pm 0.77$ & This work (Appendix~\ref{appendixd2g}) \\
Clump masses corrected by Eq.~\ref{gtdelia} & $\sfreliacorr \pm \esfreliacorr$  & $1.65 \pm 0.61$ & This work  \\
Whole 70~\micron\ emission & \multicolumn{2}{c}{$\sfrtotpacs^{+\esfrtotpacsdexplus}_{-\esfrtotpacsdexminus}$} & This work \\
EKO with new distances & 1.47 & & This work \\
\hline
Reference value & \multicolumn{2}{c}{$1.65 \pm 0.19$} & \citet{2015ApJ...806...96L} \\
\hline
\end{tabular}
\end{center}
\end{table*}

To enable the comparison among different estimates of the Milky Way SFR obtained in this paper, as well as with values in the literature, they are reported in Table~\ref{sfrsummary}.

\subsection{SFR of the Galaxy as a function of Galactocentric radius}

\begin{figure}[h!]
\includegraphics[width=0.5\textwidth]{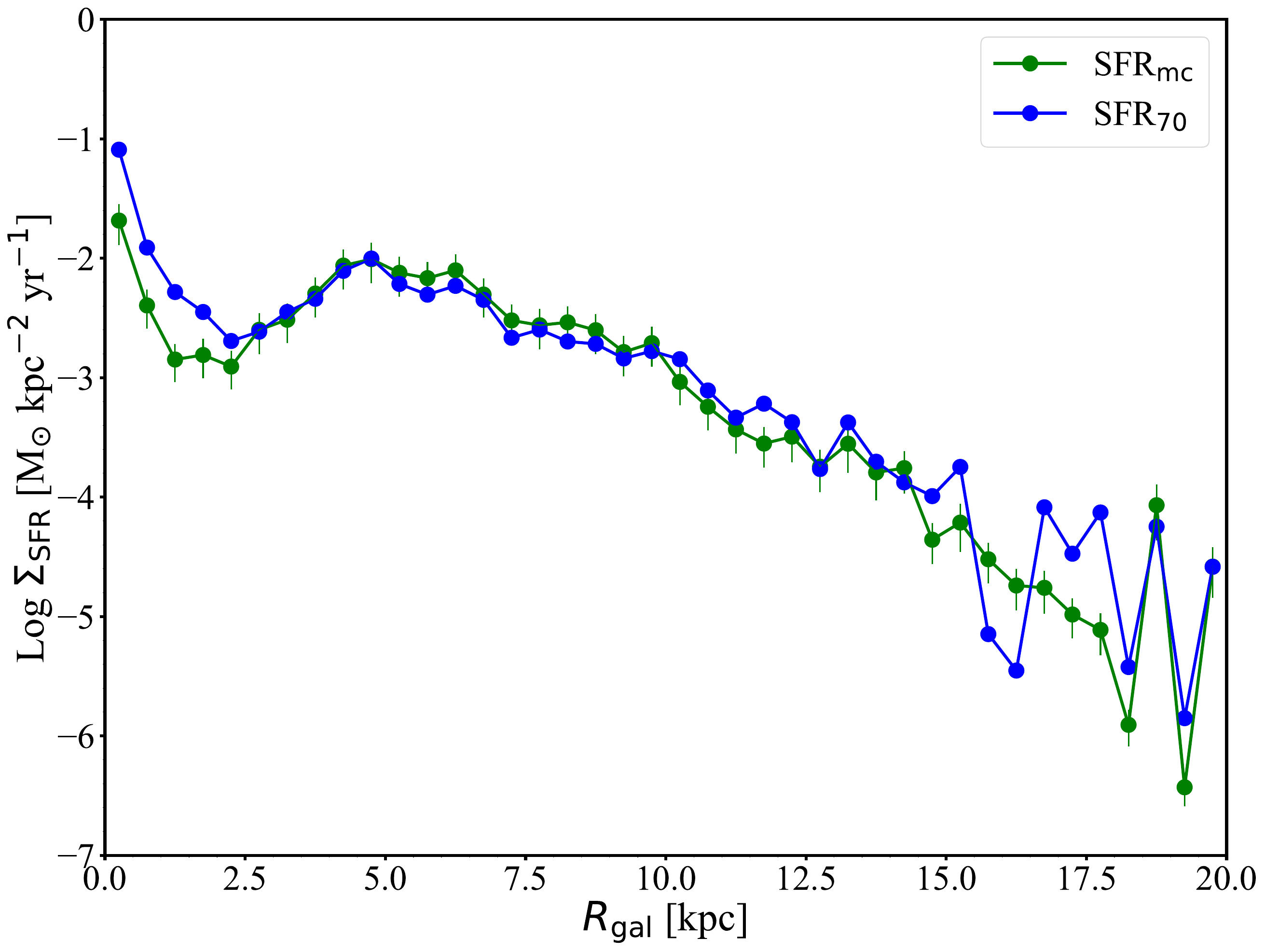}
\caption{
Comparison of SFR obtained from Hi-GAL data binned in, and plotted versus, \rgal\ (blue). 
The green points are the SFR calculated from the model of clump star formation in \citet{2022ApJ...941..162E}, but corrected for a gas-to-dust ratio given by Eq.~\ref{gtdelia}. The distributions are both normalized to a total star formation rate of 1.65~\msunyr\ \citep{2015ApJ...806...96L}, so that only the shapes can be compared.}
\label{sfr70profile}
\end{figure}

A common characteristic of the two methods, the one using the 70~\micron\ emission and the one using the clump masses, is that both are based on summing the contributions from ``basic'' elements (which is the pixel intensity at 70~\micron\ for the former, and the clump mass in the latter). This feature allows us to map the SFR density throughout the Galactic plane, and to obtain Galactocentric profiles for it. In Figure \ref{sfr70profile}, the SFR surface density (\sigmasfr) is plotted versus \rgal\ for both the 70~\micron\ method and the method based on clump masses \citep{2022ApJ...941..162E}, using the results in \S \ref{gtd}. The two estimates agree remarkably well, especially for $3 < \rgal < 8$~kpc. Both are normalized to a total SFR of $1.65$~\msunyr, so that only the shapes are compared. Inside $\rgal = 3$~kpc, the estimate from 70~\micron\ is significantly larger. Because both are normalized to the same total \sfr, the differences at very small and large \rgal\ are compensated by slightly lower predictions from $\sfrseventy$ at intermediate \rgal. 
Beyond $\sim 10$~kpc, we are outside the range of star formation rate surface density calibrated by \citet{2010ApJ...725..677L} of $\Sigma_\mathrm{SFR} > 1\ee{-3}$~\msunyr~kpc$^{-2}$, and below the metallicity limit, so we are extrapolating in both.
Nonetheless, the two methods agree rather well to at least 14~kpc, suggesting that the 70~\micron\ method may work on average even to lower flux levels. Even beyond 14~kpc, there is some agreement on average despite the small number statistics.

\subsection{SFR of the Galaxy in Two Dimensions}\label{sfr2d}

\begin{figure*}[ht!]
\includegraphics[width=\textwidth]{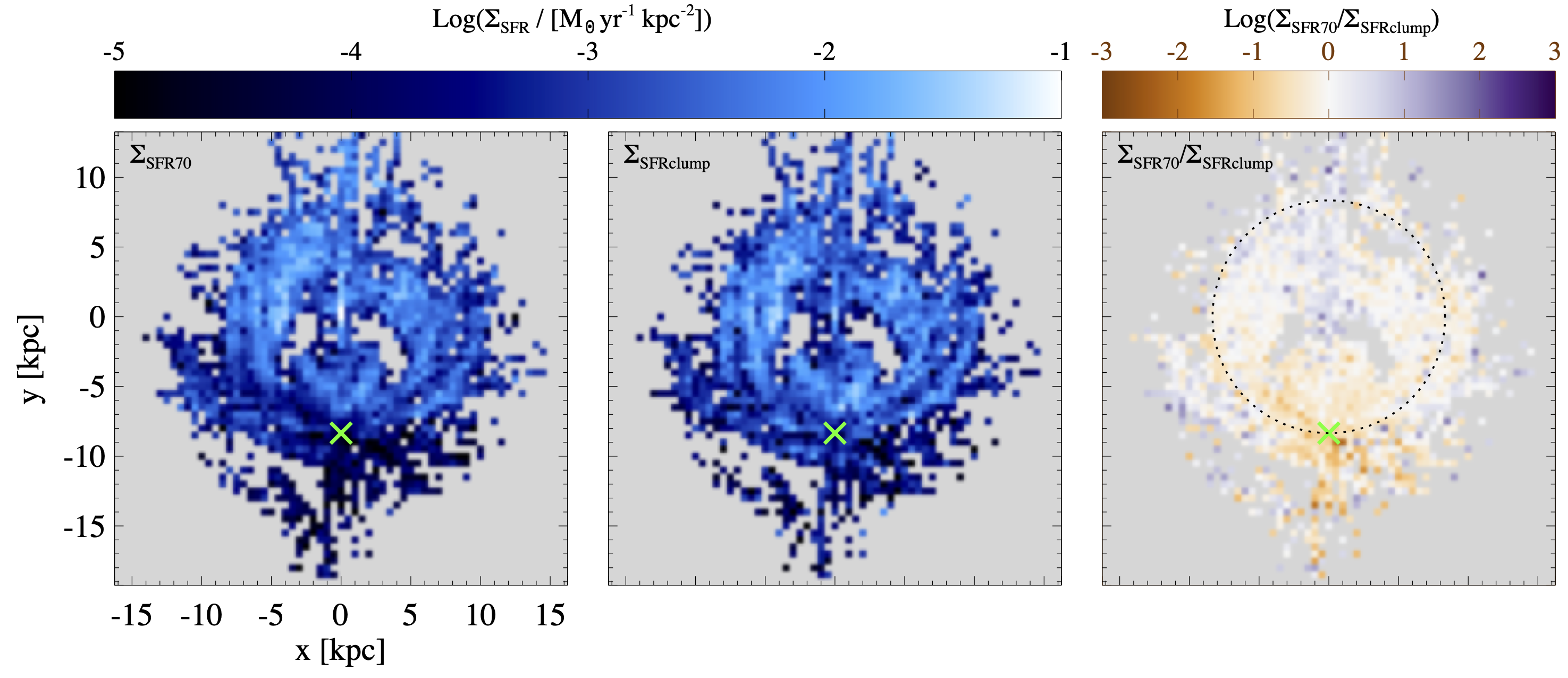}
\caption{Face-on mapping of the Milky Way SFR surface density (logarithm). \textit{Left}: map obtained with the method involving the whole \textit{Herschel} 70~\micron\ emission. \textit{Middle}: map obtained from the method considering \textit{Herschel} clump masses, corrected through Eq.~\ref{gtdelia}.
For this panel and for the previous one, a common color bar is displayed on the top. \textit{Right}: logarithm of the ratio of the two maps shown in the previous panels. The Sun position is marked with a green cross in all panels, while the Solar circle is displayed in the right panel as a dotted line).}
\label{sfrmapxy}
\end{figure*}

Again thanks to the parcelling inherent in the SFR calculation method, we can obtain 2D maps of the SFR surface density (\sigmasfr). In the left panel of Figure~\ref{sfrmapxy} the SFR surface density obtained through 70~\micron\ intensities, binned in $0.5 \times 0.5$~kpc boxes, is shown. Qualitatively speaking, it is not much different from that obtained with the clump mass method by \citet{2022ApJ...941..162E}, shown in the middle panel of Figure~\ref{sfrmapxy}. This is not surprising, since the two methods share the same set of distances and longitudes (hence Galactic locations), those of the star forming (70-\micron\ bright) clumps of \citet{2022ApJ...941..162E}. What clearly differentiates the two, however, is the independent and complementary way of calculating the SFR in those positions: in the former case by integrating the emission at 70~\micron\ from all the pixels surrounding the clump, while in the latter by considering the clump mass obtained from the fit of the portion of the SED at $\lambda\geq 160$~\micron, so not involving at all the 70~\micron\ photometry. In this respect, the observed overall similarity of the maps further supports the general consistency of the two methods discussed up to this point. Minor differences can be noticed in the first two panels of Figure~\ref{sfrmapxy}, and they are expressly highlighted in the right panel, showing the logarithm of the ratio of the two maps. On average, the 70~\micron-based method gives higher SFR surface densities in some locations in the outer Galaxy, while the inner Galaxy and the portion of the third quadrant closer to the anti-centre are dominated by the prominence of the output of the clump mass-based method. 


\subsection{Caveats}\label{caveats}

While the agreement between \sfrseventy\ and \sfrmc\ in Figure~\ref{sfr70profile} is remarkably good, we point out a few differences. The predictions of $\sfrseventy$ inside about 3~kpc exceed those from \sfrmc. A similar trend may be visible in the outer Galaxy beyond $\sim 12$~kpc, though the data are noisy there. Because 70~\micron\ emission can arise from dust in the diffuse ISM, heated by older stars
\citep{2012ARA&A..50..531K, 2010ApJ...725..677L}, 
over-prediction of SFRs can be expected in regions of low star formation rate. The outer Galaxy is clearly one such region, and the part of the Galaxy inside 4~kpc, which is affected by the stellar bar \citep{2005ApJ...630L.149B}, is another.

On the other hand, $\sfrseventy$ will underestimate the star formation rate in regions where the IMF is not well sampled 
\citep{2014MNRAS.438.2355D}. 
That will be the case in clouds and clumps of low mass, which are more easily detected by Hi-GAL when they are nearby. This effect can be seen in the orange tint in the right panel of Figure \ref{sfrmapxy} for a region a few kpc in radius around the Sun. This problem is not apparent in Figure \ref{sfr70profile} because the more distant parts of the Galaxy overwhelm the local regions in the average for $\rgal \sim 8$~kpc.

Our choice for the dependence of the gas-to-dust ratio on \rgal\ is not well established, and may require revision in the future. In particular, the trend in metallicity and thus gas-to-dust ratio inside $\rgal = 4$~kpc is uncertain \citep{2022MNRAS.510.4436M, 2024ApJ...973...89P}. In addition, the choice of 100 for the gas-to-dust ratio for the Solar neighborhood metallicity may be too low for very distant clumps, where Hi-GAL samples more diffuse cloud material, but ices do accumulate by $\av = 3.2$~mag \citep{2013ApJ...774..102W}. Variations with extinction thresholds are probably less than those caused by the primary metallicity dependence.

\subsection{SFR of the Galaxy Compared to Methods using Stars}

\begin{figure}[h!]
\includegraphics[width=0.5\textwidth]{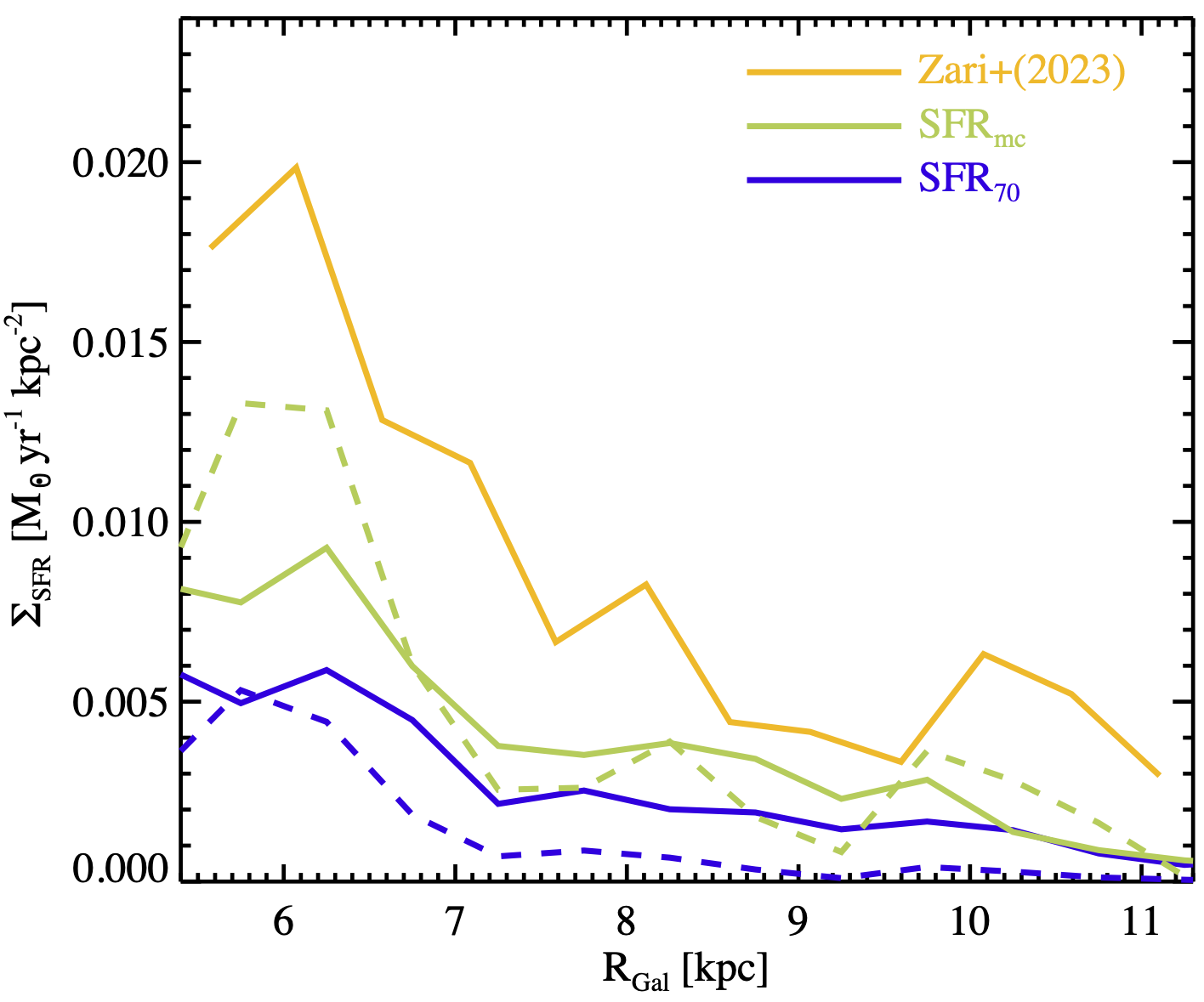}
\caption{Comparison between profiles of SFR density. The yellow curve represents that obtained by \citet{2023A&A...669A..10Z}, in bins of $\rgal=0.5$~kpc, from modeling of the populations of high-mass stars in the age interval $5 < \tau < 10$~Myr in a 6~kpc~$\times$~6~kpc portion of the Galactic plane centered on the Sun. The green solid line is obtained from clump masses (Eq.~\ref{sfrclump}) corrected by Eq.~\ref{gtdelia}, and the blue solid curve is the $\Sigma_\mathrm{SFR}$ profile obtained from 70~\micron\ emission. The green and blue dashed lines are computed as are the solid lines of the same color (see above), but considering only the 6~kpc~$\times$~6~kpc box of \citet{2023A&A...669A..10Z}.}
\label{sfroba}
\end{figure}

A direct comparison can be made, in terms of Galactocentric profile, between the SFR density obtained with the two techniques adopted in this paper and the one elaborated by \citet{2023A&A...669A..10Z}. These authors produced a map of the stellar age distribution across a 6~kpc~$\times$~6~kpc portion of the Galactic disk centered on the Sun by using a sample of $\sim 500000$ candidate O-, B-, and A-type (hereafter OBA) stars. They were able to obtain a total SFR \citep[see][for details]{2023A&A...669A..10Z}. \citet{2023A&A...678A..95S} compared the radial $\Sigma_\mathrm{SFR}$ profiles of \citet{2023A&A...669A..10Z} and \citet{2022ApJ...941..162E}, highlighting the similarity of the two in qualitative terms. They also suggested explanations for the fact that the method using OBA stars predicts a higher SFR than the clump mass method at each \rgal. Similarly to them, in Figure~\ref{sfroba} we provide an updated comparison between the data of \citet{2023A&A...669A..10Z} and those obtained in this paper. The SFR density profile obtained from clump masses remains qualitatively similar after the correction given by Eq.~\ref{gtdelia}, but it is still lower than that derived from OBA stars, as in \citet{2023A&A...678A..95S}. The discrepancy is larger if only clumps in the 6~kpc~$\times$~6~kpc box centered on the Sun are considered, with the exception of radii around 6~kpc, where it exhibits a bump which is similar (although lower) than that seen in the profile of \citet{2023A&A...669A..10Z}. Finally, the SFR density profile obtained from the 70~\micron\ maps appears to have a similar decreasing trend with \rgal\ and is lower than \sfrmc. The prediction of $\sfrseventy$ is even lower if restricted to the local 6~kpc~$\times$~6~kpc region.
While the predictions over the whole Galaxy and restricted to the local region are similar for \sfrmc, they differ substantially for $\sfrseventy$, with the local region below the Galactic average for these radii. This is presumably caused by the fact that Hi-GAL picks up lower mass clumps in the local region, which do not completely sample the IMF, as discussed in \S \ref{caveats}.

\subsection{Comparison to Method Using CO Cloud Catalogs and Theory}

The star formation surface density versus \rgal\ can also be compared to that predicted by \citet{2022ApJ...929L..18E}, hereafter denoted EKO. This prediction used the cloud catalog from
\citet{MD17},
along with some theoretical predictions described below, a very different set of inputs from either of the methods described above. 

For the sake of consistency, the cloud heliocentric distances of 
\citet{MD17}
were recalculated, starting from the velocity with respect to the local standard of rest (\vlsr) and Galactic longitude they quoted, but by applying the same rotation curve adopted by 
{\citet{2021A&A...646A..74M}
for the Hi-GAL clump distances, namely that of 
\citet{2017A&A...601L...5R}.
We kept the same solution of the near/far distance ambiguity proposed by 
\citet{MD17}, except for those far distances implying a violation of the $|z| < 140$~pc condition we imposed in \S \ref{compother}, for which we chose the near distance. Finally, for 11 sources close to the Galactic center for which there is incompatibility between the longitude, \vlsr\ pair and the rotation curve of 
\citet{2017A&A...601L...5R}, we adopt the distance quoted by 
\citet{MD17}, decreased by 0.16~kpc, namely the difference between the values of $R_0$ used in the two rotation curves. The cloud sizes and masses were rescaled to the newly calculated distances. Then, the masses were adjusted as a function of \rgal, using the same formula for the luminosity-to-mass conversion factor favored by EKO:
\begin{equation}
\alphaco = 4.50\, Z^{-0.80},
\end{equation}
where $Z$ is the metallicity in solar units, with a radial gradient of $-0.044$ dex/kpc, and \alphaco\ has the units, \alphacounit.
Densities and free-fall times were calculated from masses and sizes, and Eq.~9 from EKO (with a core-to-star efficiency of 0.30)
\begin{equation}\label{koeqn}
\epsilon_{\rm ff} = 0.30\, \exp(-2.018\, \alphavir^{1/2}),
\end{equation} 
was used to calculate \epsff, the star formation efficiency per free-fall time, and finally the SFR for each cloud in the catalog of \citet{MD17}. A new catalog, based on the catalog of \citet{MD17}, but with the distances used here, along with updated cloud properties and the calculated SFRs, is provided and described in Appendix \ref{newcat}.

The resulting star formation rate for the entire Galaxy, SFR$_\mathrm{EKO}$, is 1.47~\msunyr, again in remarkable agreement with the observational values (see 
Table \ref{sfrsummary}). While the uncertainties are mostly systematic and hard to quantify, we estimated some uncertainties for the radial distribution based simply on the number of clouds in a bin. For the largest radius plotted ($\rgal = 19.5$~kpc), the number of clouds in a bin is only 6. Comparing to figures in EKO, the change to the new distances made no discernible difference in the radial dependence and the total SFR changed only by 0.01~\msunyr.

\begin{figure}[h!]
\includegraphics[width=0.5\textwidth]{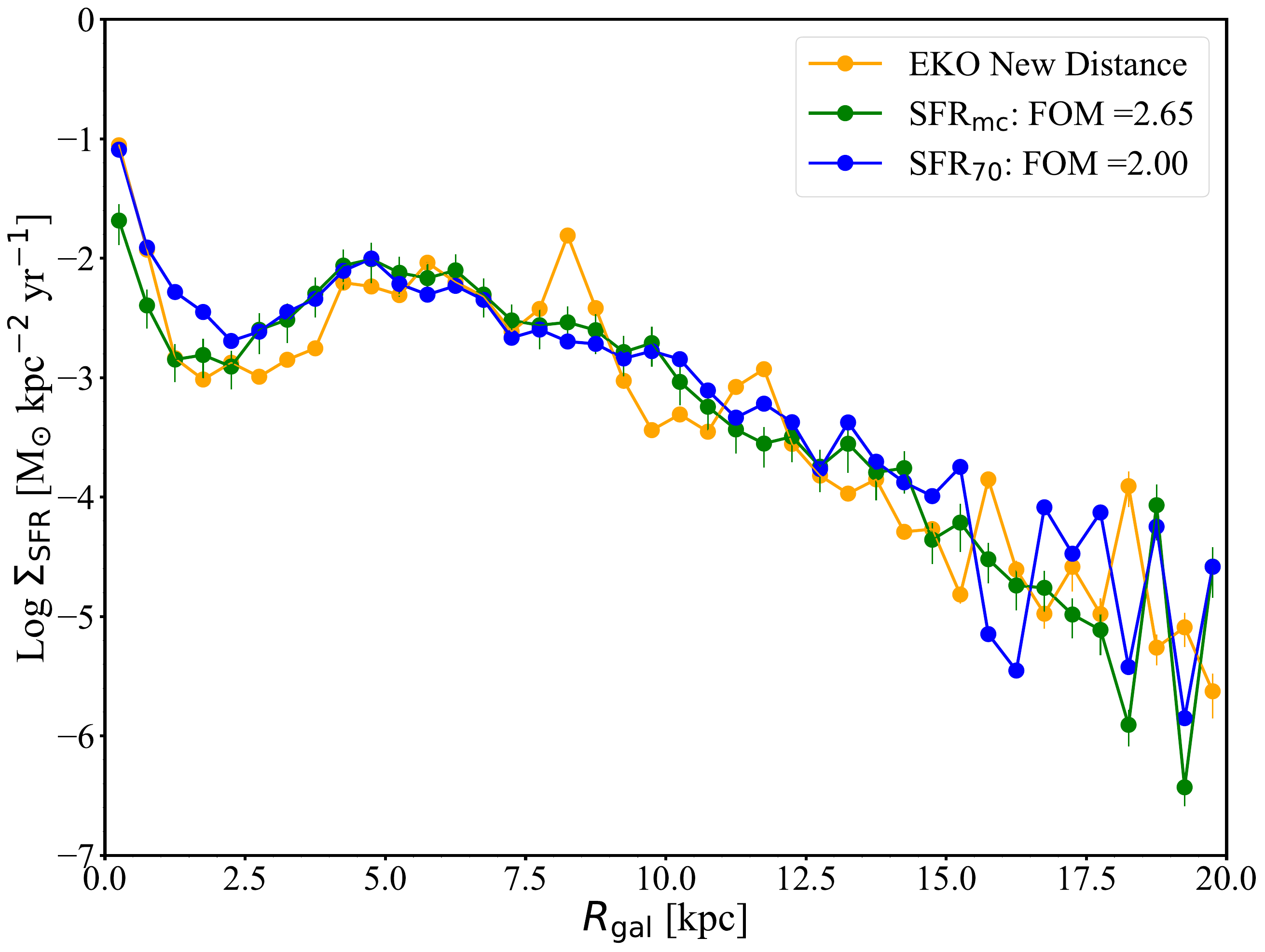}
\caption{Comparison of the SFR obtained from a EKO prediction \citep{2022ApJ...929L..18E} based on a molecular cloud catalog and an efficiency per free-fall time (orange) with those
obtained from the 70~\micron\ emission in Hi-GAL tiles binned in and plotted versus \rgal\ (blue) and 
the SFR (green) calculated from the model of clump star formation in \citet{2022ApJ...941..162E}. 
The EKO predictions are labeled EKO New Distance to indicate that they are based on the new distance estimates. The distributions are all normalized to a total star formation rate of 1.65~\msunyr\ \citep{2015ApJ...806...96L}, so that only the shapes are compared.}
\label{sfr70vmod}
\end{figure}

Both observational methods (the 70~\micron\ and clump mass) show excellent agreement with the EKO prediction of the radial distribution of SFR (Figure \ref{sfr70vmod}), all normalized to 1.65~\msunyr\ total to focus on the radial dependence. The \sfr\ based on 70~\micron\ emission matches the EKO prediction slightly better, as reflected in the slightly smaller figure of merit, defined by $\rm{FOM} = \mean{|{\rm (obs/EKO)}- 1|}$. In particular, the match in the innermost \rgal\ makes the two lines indistinguishable in the figure. The clump mass model agrees better with the EKO predictions around 2~kpc, where both predict a lower star formation rate. The most notable disagreement occurs around 8~kpc, where the EKO method predicts a jump of about a factor of 3 that is not seen in either of the observational distributions, perhaps influenced by local clouds\footnote{In general, the SFR estimates based on Hi-GAL data may be affected by the limited coverage in latitude (a strip of $\Delta b=2^\circ$) of this survey, which excludes the contributions to SFR from high-$|b|$ star forming regions. However, the impact of this effect appears to be relatively minor in the context of the entire Galactic plane, as it involves a limited number of clouds at a relatively small $d$. For example, \citet{2010ApJ...724..687L} report the SFRs of 11 nearby star forming regions, with the highest being that of Orion~A at 7\ee{-4}~\msunyr; the combined SFR of all 11 regions does not even reach twice this amount. Nonetheless, this effect may not be negligible when estimating SFR on more localized scales and could contribute to the observed discrepancy between the two radial SFR density profiles highlighted here, because the cloud catalog of \citet{MD17} covers $|b| \leq 5^\circ$.\label{footnote_local}}. However, given uncertainties, both agree well with the EKO predictions. This agreement is rather remarkable because the EKO predictions were based on a completely different data base and method. The cloud catalog was based on CO \jj10\ data with lower spatial resolution \citep{Dame01} and a different cloud identification algorithm \citep{MD17}. 
The agreement between all of these methods gives support to each.

\begin{figure*}[ht!]
\includegraphics[width=\textwidth]{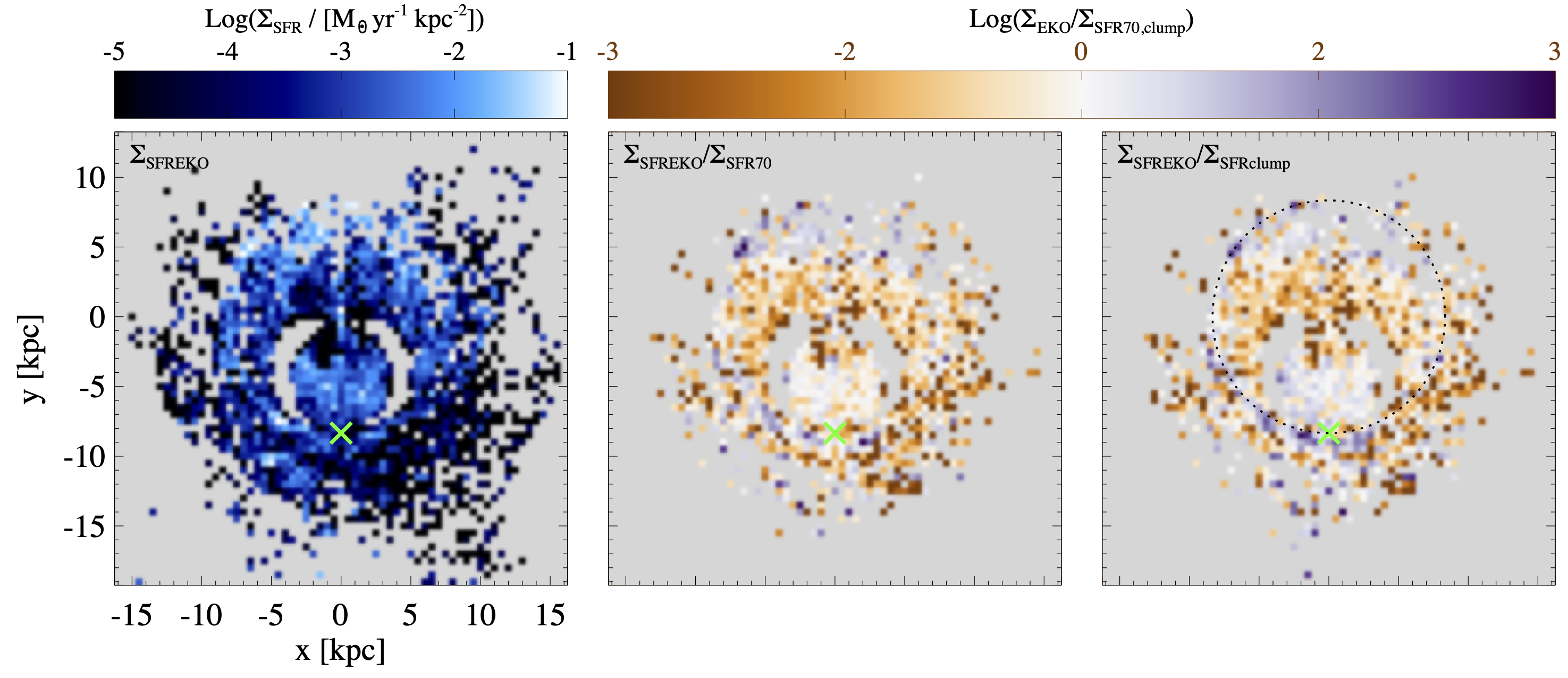}
\caption{\textit{Left}: face-on mapping of the Milky Way SFR surface density (logarithm) using the EKO method and the CO clouds cataloged by \citet{MD17}, with the distances assigned in this paper. The color bar on the top shows the range of the mapped values. \textit{Middle}: logarithm of the ratio between the map in the left panel of this figure and the one in the left panel of Figure~\ref{sfrmapxy} (SFR map obtained from the whole \textit{Herschel} 70~\micron\ emission). \textit{Right}: logarithm of the ratio between the map in the left panel of this figure and the one in the middle panel of Figure~\ref{sfrmapxy} (SFR map obtained from Hi-GAL clump masses). The middle and right panel share a common color bar displayed on the top. The Sun position is marked with a green cross, while the Solar circle is displayed in the right panel with a dotted line.}
\label{sfrmapmiv}
\end{figure*}

The CO-traced clouds cataloged by \citet{MD17} can be also used, after conversion to the same model of Galactic rotation, to obtain the corresponding face-on $\Sigma_\mathrm{SFR}$ map (Figure~\ref{sfrmapmiv}, left), analogous to the Herschel-based maps shown in Figure~\ref{sfrmapxy}. 
A comparison of this map with the $\Sigma_\mathrm{SFR}$ map obtained from the diffuse 70~\micron\ emission is shown in the form of pixel-to-pixel ratio in the middle panel of Figure~\ref{sfrmapmiv}. A trend to contain a less prominent spiral-arm structure and to show a higher $\Sigma_\mathrm{SFR}$ in the Solar vicinity appears in the CO cloud-based map. This trend is further highlighted in the comparison of this map with that obtained from clump masses (Figure~\ref{sfrmapmiv}, right panel). The detailed agreements of this map with the other two are less impressive than the agreement of the clump mass and 70~\micron\ methods. SFR estimates may differ substantially in individual regions while agreeing when averaged over sufficiently large scales.
While the radial distributions predicted by the EKO model agree very well with the clump mass and 70~\micron\ methods, the face-on maps show substantial disagreements within individual 0.5~kpc boxes (Figure \ref{sfrmapmiv}). Disagreements are especially common at large \rgal, where small numbers of clouds and clumps limit the benefits of averaging.

\subsection{Star Formation Relations}

\begin{figure}[h!]
\includegraphics[width=0.5\textwidth]{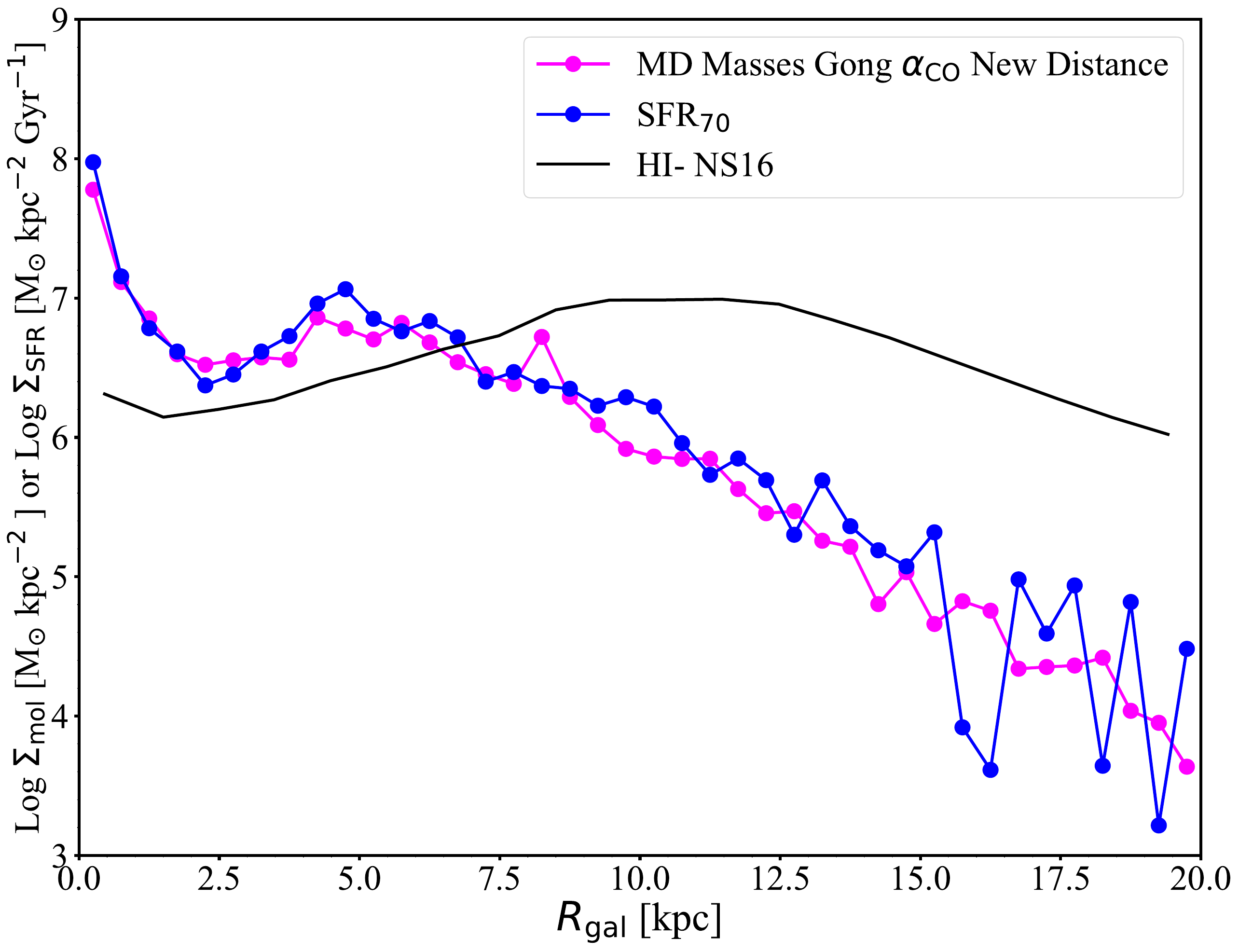}
\caption{The radial distribution of surface densities of atomic gas \citep[black, from][]{2016PASJ...68....5N}, molecular gas \citep[magenta, from][but after distance correction carried out in this work]{MD17}, and SFR derived in this work through the 70~\micron\ method (blue).}
\label{keupdate}
\end{figure}

The excellent agreement of three methods to obtain the radial distribution of the star formation rate provides the opportunity to update Figure~7 of \citet{2012ARA&A..50..531K}. We computed the radial dependence of the molecular cloud mass surface density ($\Sigma_{\rm mol}$) from the new catalog, using the same rotation curve as the Hi-GAL project used and $\alphaco(Z)$. The result is plotted in Figure~\ref{keupdate}, along with the $\Sigma_{\rm SFR}$ based on the 70~\micron\ method, as this is the estimate of SFR most based on pure observational data. Unlike Figure~7 of \citet{2012ARA&A..50..531K}, the data for both molecules and star formation are now much better determined and agree extremely well over the full range of \rgal\ from the CMZ to the far outer Galaxy. By plotting the SFR per Gyr on the $y$-axis, we can immediately see the depletion time ($\tdep = \mmol/\sfrseventy$) variation in any offset between it and the mass distribution. The good agreement shows that the depletion time for molecular gas in the Milky Way  is about 1~Gyr at all \rgal. More precisely, the total molecular mass within 20~kpc is 1.34\ee9~\msun. With a star formation rate from the 70~\micron\ emission of 1.42~\msunyr, the depletion time is 0.94 Gyr, about half the average for nearby galaxies \citep{2011ApJ...730L..13B, 2024ARA&A..62..369S}. The surface density of atomic gas ($\Sigma_{\rm atomic}$) from \citet{2016PASJ...68....5N} has a completely different and relatively flat distribution, as also highlighted by \citet{2022ApJ...941..162E}.

The plots of $\log(\Sigma_{\rm SFR})$ versus \rgal\ show a mostly linear behavior for \rgal\ well beyond the radii affected by the bar, indicating an exponential model: $\Sigma_{\rm SFR} \propto \exp(-\rgal/a)$, where $a$ is the scale length. However, fits restricted to ranges of \rgal\ show that the slope steepens beyond $\rgal \sim 14$~kpc. Measurements of the scale length for the stellar disk vary substantially, but \citet{2016ARA&A..54..529B} give an average value of $a = 2.6 \pm 0.5$~kpc. Recent studies of the stellar distribution favor a flat distribution, a linear region, and a steepening \citep{2024NatAs.tmp..190L}, also beyond 14~kpc. They show that the distribution of all stars between 7.5 and 14~kpc agrees with an exponential of scale length 2.6~kpc. Because the SFR density does not show the flat region (affected by the stellar bulge), we fit linear slopes between 5.0 and 14~kpc for the best comparison to the stars.
These fits lead to the following scale lengths: for the clump mass model, $a = (2.03 \pm 0.10)$~kpc; 
for the 70~\micron\ method, $a = (2.47 \pm 0.16)$~kpc; 
for the EKO method, $a = (1.94 \pm 0.26)$~kpc.
The average of all three yields $\mean{a} = (2.15 \pm 0.28)$~kpc, marginally shorter than the stellar scale length. \citet{2022ApJ...941..162E}, using the clump mass method but with uncorrected masses, also found a decreasing exponential trend, steeper than the aforementioned ones (see Appendix~\ref{appendixd2g}), and whose slope can be translated in terms of the scale length used here, as $(1.55 \pm 0.06)$~kpc. 
These can be compared to the fit for smoothed surface density of molecular gas ($\Sigma_{\rm mol}$), with a 2.0~kpc scale length (Fig.~9 of \citealt{MD17}). With the new distances and CO luminosity to mass conversion factor that varies with $Z$, the scale length of $\Sigma_{\rm mol}$ is $(2.26 \pm 0.14)$~kpc.

\begin{figure}[h!]
\includegraphics[width=0.5\textwidth]{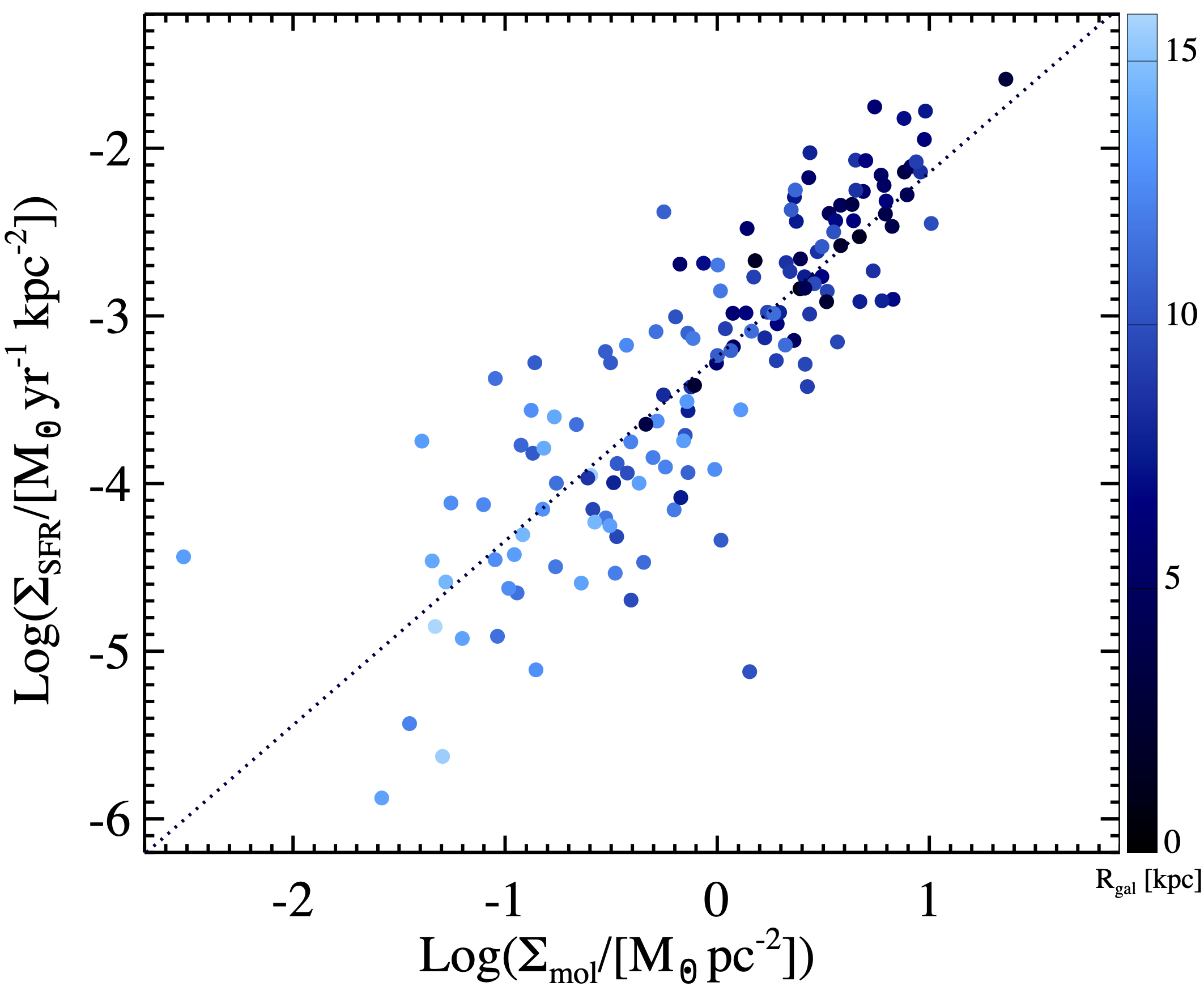}
\caption{SFR surface density calculated through the 70~\micron\ emission method (Figure~\ref{sfrmapxy}, left panel), averaged in $2 \times 2$~kpc$^2$ bins, versus the smoothed surface density of molecular gas, from \citet{MD17}, averaged in the same bins. 
The colors of the symbols are encoded based on the Galactocentric distance, as indicated by the color scale shown in the bar on the right.
The dotted line represents the power-law fit to data, highlighting a Kennicutt-Schmidt behavior of the plot, with power-law exponent $n=1.10\pm 0.06$.}
\label{ksplot}
\end{figure}

Since both $\Sigma_{\rm SFR}$ and $\Sigma_{\rm mol}$ show an exponential behavior with respect to \rgal, a power-law is expected between the two, which corresponds to the Kennicutt-Schmidt (KS) relation \citep{1959ApJ...129..243S,1998ApJ...498..541K} applied to molecular gas: $\Sigma_{\rm SFR} \propto \Sigma_{\rm mol}^n$. 
In Figure~\ref{ksplot} the plot of $\Sigma_{\rm SFR}$, derived through the 70~\micron\ emission method, versus $\Sigma_{\rm mol}$ is shown, where $\Sigma_{\rm mol}$ is the surface density averaged over a 2~kpc square box. 
A power-law fit to the data returns a slope $n=1.10\pm 0.06$,
a nearly linear dependence of SFR on molecular cloud mass, suggested by studies in other galaxies \citep{2011ApJ...730L..13B}.} 
In the outer Galaxy, we probe $\Sigma_{\rm mol}$ as low as $\sim 2\ee{-2}$~\msunpc, more than an order of magnitude lower than in Figure~1 of \citet{2011ApJ...730L..13B}, allowing an extension (albeit noisy) in the range of the KS relation. 
We emphasize that the quantity represented on the $x$-axis in Figure \ref{ksplot} is not the average surface density of a cloud, but the smoothed-out mass surface density of clouds obtained by dividing the sum of masses of clouds lying within a spatial bin by its area, in this case 4~kpc$^2$.


The level of scatter in the plot in Figure~\ref{ksplot} surely depends on the physical size adopted for the spatial bins, as the effect of local fluctuations is expected to be attenuated by a coarser binning. Indeed, doubling for instance the bin size, we obtain a globally narrower distribution of points around the best-fit power law, whose exponent remains $n=1.10$. Contrarily, halving the bin size creates more scatter, dominated by local fluctuations, in which an overall increasing trend is still seen, but not enough to describe it as a power-law behavior.

\subsection{Why does it work so well?}\label{sec:theory}

Both the EKO method, based on molecular clouds, and the clump mass method, based on dense gas, predict the total and radial distribution of the star formation rate based on the 70~\micron\ emission remarkably well, even for very small \rgal. In this subsection, we consider some aspects of this agreement, centered on three questions.

Why do we correctly predict the SFR in the CMZ?
The innermost point of Figures \ref{sfr70vmod} and \ref{keupdate} includes the inner 500 pc, one definition of the Central Molecular Zone (CMZ). This point is consistent with the overall exponential decrease in star formation rate defined by the data for $\rgal > 5$ kpc. In addition, Figure \ref{ksplot} indicates no departure from the overall KS relation for the smallest \rgal. 
This agreement seems contrary to expectations that the SFR is anomalously low in the CMZ. 
In fact, the value of \sfrseventy(CMZ) is $0.064 \pm 0.001$~\msunyr, completely consistent with the average value favored by
\citet{2023ASPC..534...83H}
of  $0.07^{+0.08}_{-0.02}$~\msunyr, which we take as the standard value for further comparisons.
The EKO method
\citep{2022ApJ...929L..18E}
predicts $0.062 \pm 0.006$~\msunyr, in excellent agreement. Previous work indicated that the SFR was consistent with predictions from molecular gas from studies of other galaxies 
\citep{2014MNRAS.440.3370K}.
The claim of low SFR in the CMZ is based on the dense gas relations 
\citep{2004ApJ...606..271G, 2010ApJ...723.1019H, 2010ApJ...724..687L, 2012ApJ...745..190L}.
In this context, the prediction of the clump mass model is the most interesting. In fact, our prediction is $\sfrmc = 0.033 \pm 0.013$~\msunyr, {\it lower} than the \textcolor{magenta}{value of \citet{2023ASPC..534...83H}} by two standard deviations. Thus, we find no support for the common idea that the SFR in the CMZ is low per amount of dense gas, {\it as measured by the masses of Hi-GAL compact sources, with our adopted correction for gas-to-dust ratio}. 

In fact, \citet{2022ApJ...941..162E} already found  substantial agreement between the results from the clump mass method and previous estimates of the SFR in the CMZ from the literature \citep{2011MNRAS.413..763C,2012A&A...537A.121I,2017MNRAS.469.2263B}. The only exceptions were the findings of \citet{2013MNRAS.429..987L}, who underestimated the SFR, and of \citet{2009ApJ...702..178Y}, who overestimated it. For these two cases, \citet{2022ApJ...941..162E} provided well-supported justifications. We recalculated the SFR using the methods introduced here, based on the whole 70~\micron\ emission and clump masses corrected for variable gas-to-dust ratio, respectively, within the $[\ell,b]$ boundaries of the sky areas investigated by those authors. The results are quantitatively consistent with those of \citet{2022ApJ...941..162E}, therefore confirming their conclusions.

The original CMZ puzzle began with 
\citet{2013MNRAS.429..987L},
who estimated that there were $\eten7$~\msun\ of dense gas in the CMZ from strong, extended ammonia emission. In contrast, the catalog of 
\citet{2021MNRAS.504.2742E}
 finds a total of $1.56\ee6$~\msun\ in dense clumps (both prestellar and protostellar) within $\rgal = 0.5$~kpc. If we correct for the lower gas-to-dust ratio in the CMZ (39 at $\rgal = 0.25$~kpc), this mass drops to $6.03\ee5$~\msun. The 
\citet{2012ApJ...745..190L}
relation would then predict a star formation rate of $0.028$~\msunyr, in agreement with the prediction by the clump mass method. The issue comes down to the definition of ``dense''.  With the gas-to-dust ratio at 0.25~kpc of 39, the average surface density of clumps within 0.5~kpc is $0.36$~g~\cmc, and the average volume density is $5.0\ee4$~\cmv, considerably higher than the average density sampled by ammonia emission of $7.6\ee2$~\cmv, the effective density at $\tk = 20$~K 
\citep{2015PASP..127..299S},
which in turn is 
considerably lower than the ``several $\times 10^3$ \cmv", assumed by 
\citet{2013MNRAS.429..987L}. 
Over-estimation of the density ``traced" by so-called dense gas tracers is a common problem (see discussion in 
\citealt{2020ApJ...894..103E}),
and the mass of gas traced by these lines includes much gas at rather low densities
\citep{2020ApJ...894..103E,2022AJ....164..129P}.

Why is the prediction of SFR from the compact sources so low?
The star formation rate computed from the 70~\micron\ emission from all the compact sources in the catalog of
\citet{2021MNRAS.504.2742E}
is about 100 times less than that computed from the whole 70~\micron\ emission. This is interestingly similar to the ratio of energy injection rates from fusion and accretion by the approximate end of a star formation event in a model cluster, as seen in Figure~5 of 
\citet{2022MNRAS.512..216G}.
These ratios could be related if the accretion luminosity is captured locally, while the fusion-powered luminosity is more distributed over the clump and resulting cluster. If so, the whole 70~\micron\ emission from a clump is averaging over a longer time, as the luminosity from fusion exceeds that from accretion by 3~Myr in the simulation and reaches the factor of 100 by about 6~Myr.

Does all the diffuse 70~\micron\ emission trace recent star formation?
The assignment of all the extended 70~\micron\ emission to the nearest compact source, while necessary to assign a distance, can be challenged. Simulations of the creation and propagation of far-ultraviolet light (FUV) in large sections of galaxies indicate that it contributes about twice the dust heating as do optical photons, which can arise from older stars
\citep{2023ApJS..264...10K,2024ApJ...975..173L}. 
Contributions from diffuse 70~\micron\ emission far from the compact source can be due to heating by older stars. To check the importance of this effect, we assessed the contribution to the whole emission as a function of separation from the compact source. We find that less than 15\% of the total 70~\micron\ emission from the Milky Way arises more than 32~pc from a compact source, the average size of a CO cloud in the catalog of
\citet{MD17}.

\section{Summary}\label{summary}

In this paper, two independent techniques for estimating the Galactic star formation rate from \textit{Herschel} data have been originally elaborated and applied. We have also re-calculated the star formation rates from a third method (EKO). The total star formation rates based on the three methods are in excellent agreement (Table \ref{sfrsummary}). Our further results can be summarized as follows:
\begin{itemize}
  \item We explored the consequences of correcting Hi-GAL clump masses, used as input of the SFR calculation algorithm of \citet{2022ApJ...941..162E}, by a gas-to-dust ratio variable with \rgal. Using the profile suggested by \citet{2017A&A...606L..12G} emphasizes the contribution of the outer Galaxy at the expense of that of the inner Galaxy, producing a global SFR very close to the ``uncorrected" one, but it predicts a clump surface density increasing with \rgal\ for quiescent clumps.
 \item We instead propose a weaker dependence of the gas-to-dust ratio on \rgal, based on assuming a constant surface density of Hi-GAL clumps. By applying this gas-to-dust ratio to the SFR calculation, we obtain a profile which is quite similar to, but somewhat flatter than, the ``uncorrected" one. 
 \item Using the 70~\micron\ method by \citet{2010ApJ...725..677L} with only the compact sources in the Hi-GAL catalog to calculate the star formation rate of the Galaxy fails by two orders of magnitude. This failure arises because most of the 70~\micron\ emission is extended.
    \item Using the whole emission in maps rather than using only fluxes of compact sources, the 70~\micron\ method predicts a total star formation rate and a distribution over the Galaxy that agrees very well with the predictions of the clump mass model.
\item The Galactic star formation rate obtained with the whole 70-\micron\ emission is \sfrtotpacs~\msunyr, a factor of \lioverlicq\ less than the precise value of 1.65~\msunyr\ found by \citet{2015ApJ...806...96L}. The residual discrepancy can be interpreted as a correction to the calibration factor of \citet{2010ApJ...725..677L}. 
    \item The ``modular" approach followed for the calculation of the Galactic SFR from 70~\micron\ emission allowed us to map it across the Galactic plane, and to carry out position-to-position comparisons with the method based on the masses of the clumps by \citet{2022ApJ...941..162E}. The shapes of the Galactocentric profiles of $\Sigma_\mathrm{SFR}$ given by the two methods are very similar.
    \item The Galactocentric radial profile of $\Sigma_\mathrm{SFR}$ obtained from both methods are in very good agreement with the theoretical predictions of \citet{2022ApJ...929L..18E}, even in the outer Galaxy.
\item The agreement of all three methods provides a solid description of star formation across the Galaxy, facilitating comparison to other galaxies.
\item Regions of the Galactic plane, chosen as 2~kpc $\times$ 2~kpc boxes, follow the Kennicutt-Schmidt relation between SFR density and molecular gas density, with a slope of $1.10 \pm 0.06$.
\item The 70~\micron\ and EKO methods both predict the generally accepted value of the star formation rate in the CMZ, while the clump mass method slightly under-predicts it. We find no evidence for a deficit of star formation in the CMZ in the dense gas traced by the clumps in the Hi-GAL catalog.
\end{itemize}

\begin{acknowledgments}
We are grateful to the referee, M. Heyer, and J-G. Kim for suggestions that led to the addition of section 6.8. DE acknowledges funding from INAF Mini Grant program
“The Multi-Fractal Structure of the InterStellar Medium (MSISM)”. 
NJE thanks the Astronomy Department of the University of Texas for research support, INAF-IAPS (Roma, Italy) for hospitality during a visit, and H. Khan-Farooki, along with two anonymous cornea donors, for the gift of sight.

\end{acknowledgments}

\software{astropy
\citep{2013A&A...558A..33A,2018AJ....156..123A,2022ApJ...935..167A}
}
\appendix
\section{Radial Variation in the gas-to-dust ratio}\label{appendixd2g}

An exponential variation of $\gamma$ with \rgal\ was proposed by \citet{2017A&A...606L..12G}:
\begin{equation}\label{gianneq}
    \log(\gamma_\mathrm{G}) = 0.087\,^{+0.045}_{-0.025}\: \frac{\rgal}{\rm{kpc}} + 1.44\,^{+0.21}_{-0.45} \:,
\end{equation}
where the uncertainties are systematic. Much smaller statistical uncertainties of $\pm 0.007$ and $\pm 0.03$ were also assigned to the first and second terms, respectively.
Examples of the application of this formula are $\gamma_\mathrm{G}$ values of $\sim 28$, 146, and 556 for $\rgal =0$, 8.34 (i.e., the solar circle radius ($R_0$) assumed here to derive clump distances), and 15~kpc, respectively. The value of 146 for the Solar circle is correct for the diffuse ISM when all the gas components are included
\citep{2007ApJ...663..866D}. 
However, because of depletion of gas phase species into ices in molecular clouds
\citep{2014prpl.conf..363P}, the traditional gas to dust ratio of 100 is actually more correct for the conditions in regions that dominate the Hi-GAL catalog (Patra et al., in prep.).
\citet{2024MNRAS.528.4746U} already used Eq.~\ref{gianneq} to correct masses and surface densities of Hi-GAL clumps.

Applying Eq.~\ref{gianneq}, instead of Eq.~\ref{gtdelia}, to compute the total Milky Way's SFR as described in \S\ref{gtd}, we obtain $(\sfrgianncorr \pm \esfrgianncorr)$~\msunyr, which increases to $(2.06 \pm 0.77)$~\msunyr\ if the contribution from clumps without a distance estimate is also taken into account (see Table~\ref{sfrsummary}).

\citet{2021MNRAS.504.2742E} noticed a definite decreasing trend of the median surface density $\Sigma_\mathrm{cl}$ of Hi-GAL clumps with \rgal, at all radii for the sub-class of quiescent clumps, and starting at $\rgal \sim 2-3$~kpc, even for the star forming ones. This might be due to intrinsic variation of physical conditions from the inner to the outer Galaxy, or to the use of a constant $\gamma$ rather than one increasing with \rgal\ (as, for example, in Eq.~\ref{gianneq}), or to a combination of the two. 

\begin{figure}[ht!]
 \begin{center}
\includegraphics[width=0.5\textwidth]{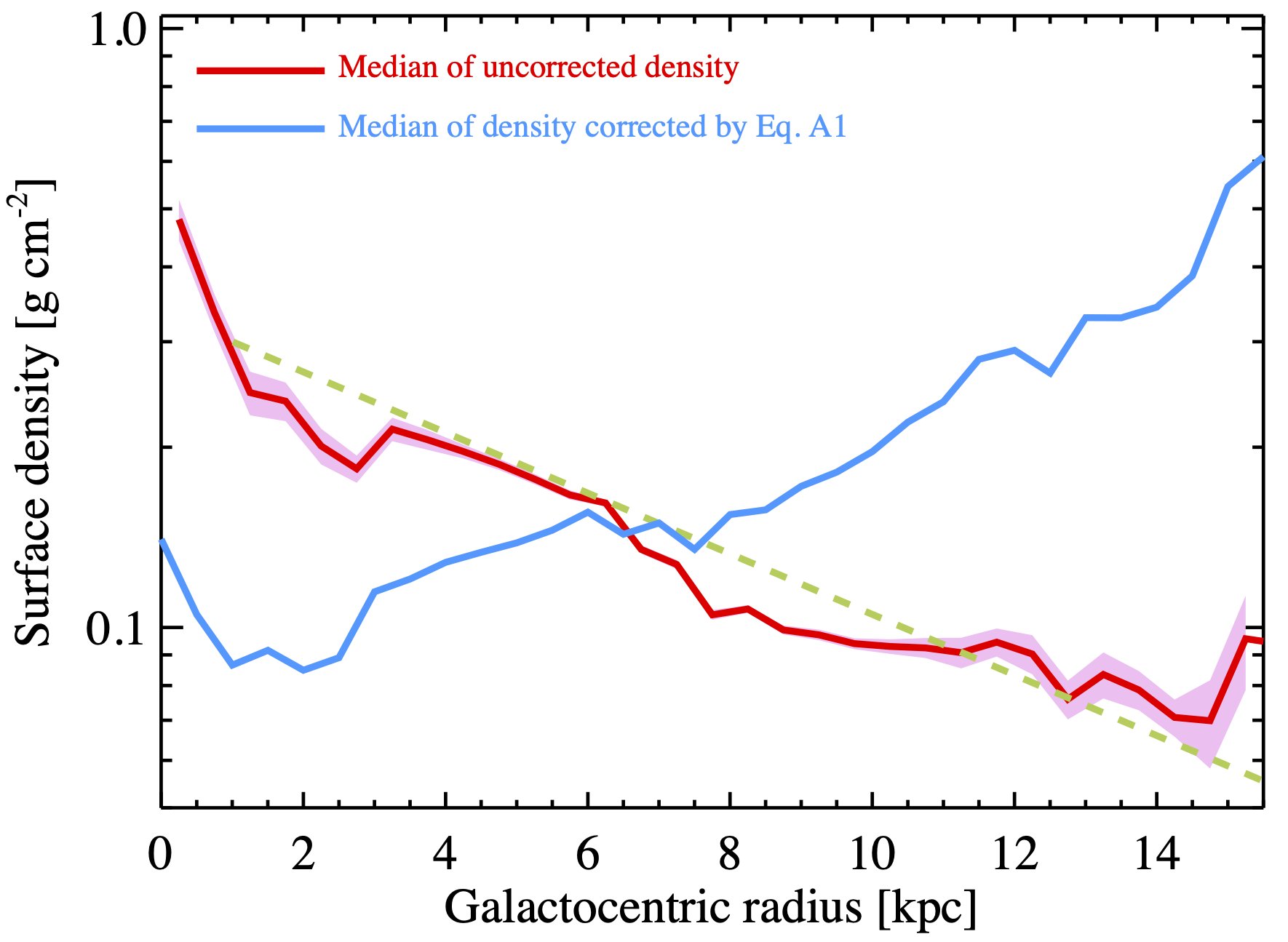}
\caption{Surface density $\Sigma_\mathrm{cl}$ vs Galactocentric distance \rgal\ of pre-stellar (i.e. quiescent and gravitationally bound) Hi-GAL clumps. The red line represents the median of $\Sigma_\mathrm{cl}$ in \rgal\ bins of 0.5~kpc, while the pink-shaded area corresponds to the uncertainty, calculated as in \citet{2022ApJ...941..162E}. The green dashed line represents the best exponential fit to the data (not to the median) for $\rgal > 1$~kpc. The blue line represents the median of $\Sigma_\mathrm{cl}$ if this quantity was rescaled by applying a gas-to-dust ratio variable with \rgal\, as suggested by \citet{2017A&A...606L..12G}. }
\label{clumpsurfd}
 \end{center}
\end{figure}

In Figure~\ref{clumpsurfd} we plot the median of $\Sigma_\mathrm{cl}$ (calculated in bins of 0.5~kpc) for the entire population of Hi-GAL quiescent clumps used in the analysis of \citet{2021MNRAS.504.2742E}. We also show how it would appear if corrected by the inverse of $\gamma_\mathrm{G}$ given in Eq.~\ref{gianneq}. This correction predicts an increase of the surface density of clumps with \rgal, which seems unlikely at large \rgal. We conclude that the $\gamma_\mathrm{G}$ given in Eq.~\ref{gianneq} has too strong a dependence on \rgal\ for the full Hi-GAL sample.

If instead we assume that the median surface density should remain nearly constant, we follow an alternative approach to constrain the gas-to-dust ratio across the whole \rgal\ range of the Galaxy. The behavior of the quiescent clumps appears to us the most suitable for this analysis. Firstly, they cover the entire range of Galactocentric distances. Secondly, they have the simplest structure in terms of temperature (having no strong internal heating source) and density, thus reducing the uncertainties connected with their modeling. Finally, they provide the highest-fidelity picture of the initial conditions of clumps (so more genuinely mirroring a possible bias introduced by the choice of a constant gas to dust ratio), before part of their mass is transferred into stars or dispersed by the star formation process.


The systematically decreasing behavior of the median of the surface density shown in Figure~\ref{clumpsurfd}
encourages us to fit an exponential law, analogously to \citet{2017A&A...606L..12G}. The best fit is found for the following combination of parameters: 
\begin{equation}
    \log\left(\frac{\Sigma_\mathrm{cl}}{\rm{g~cm}^{-2}}\right) = (-\gtdexptwo \pm \egtdexptwo)\left(\frac{\rgal}{\rm{kpc}}\right) -\gtdexpone \pm \egtdexpone \:.
\label{sigmaconsts}
\end{equation}

In this fit, we have excluded the region with $\rgal < 1$~kpc because both clouds
\citep{2001ApJ...562..348O}
and clumps
\citep{2020ApJS..249...35B}
appear to have much higher surface densities in the Central Molecular Zone, as reviewed by 
\citet{2023ASPC..534...83H}.
While corrections for metallicity effects will lower these estimates, they will not remove the full effect.
The radial gradient of $-0.051$ dex/kpc in Eq.~\ref{sigmaconsts} is similar to gradients in elemental abundances. For example \citet{2022MNRAS.510.4436M} derived a gradient in [O/H] of $-0.044 \pm 0.009$ dex/kpc based on observations of \hii\ regions. The nitrogen abundance gradient for $4 < \rgal < 17$~kpc is somewhat larger, at $-0.068 \pm 0.005$, but the [N/H] relation appears to flatten for $\rgal < 4$~kpc \citep{2024ApJ...973...89P}. The similarity of all these suggests a nearly linear dependence of the dust fraction on metallicity.

Assuming that the declining surface density of clumps in Figure \ref{clumpsurfd} is completely due to having adopted a constant $\gamma$ ratio, as described above, this result can be translated into a prescription for the gas-to-dust ratio dependence from \rgal\, expected to have the inverse behavior. Imposing the exponential law to assume the value of 100 at the Sun's Galactocentric distance $R_0 = 8.34$~kpc, we derived Eq.~\ref{gtdelia}.

Eqs.~\ref{gianneq} and~\ref{gtdelia} can be used to rescale clump masses used as the input of the SFR calculation, and obtain the corresponding Milky Way SFR estimates, which are $\mathrm{SFR}_\mathrm{G}=(\sfrgianncorr \pm \esfrgianncorr)~\msunyr$ and $\mathrm{SFR}_\mathrm{H}=(\sfreliacorr \pm \esfreliacorr)~\msunyr$, respectively. These estimates are reported also in Table~\ref{sfrsummary}, which summarizes all SFR values obtained in this paper, allowing immediate comparisons among them and with literature values. Because these global numbers are fully consistent with the value of $\mathrm{SFR}_\mathrm{mc}=(1.74 \pm 0.65)~\msunyr$ found by \citet{2022ApJ...941..162E} (involving only clumps with a heliocentric distance estimate), it is interesting, given the different analytic dependence of these two corrections, to see how the Galactocentric profile varies for these two ways of deriving SFRs.

 \begin{figure}[ht!]
 \begin{center}
\includegraphics[width=0.5\textwidth]{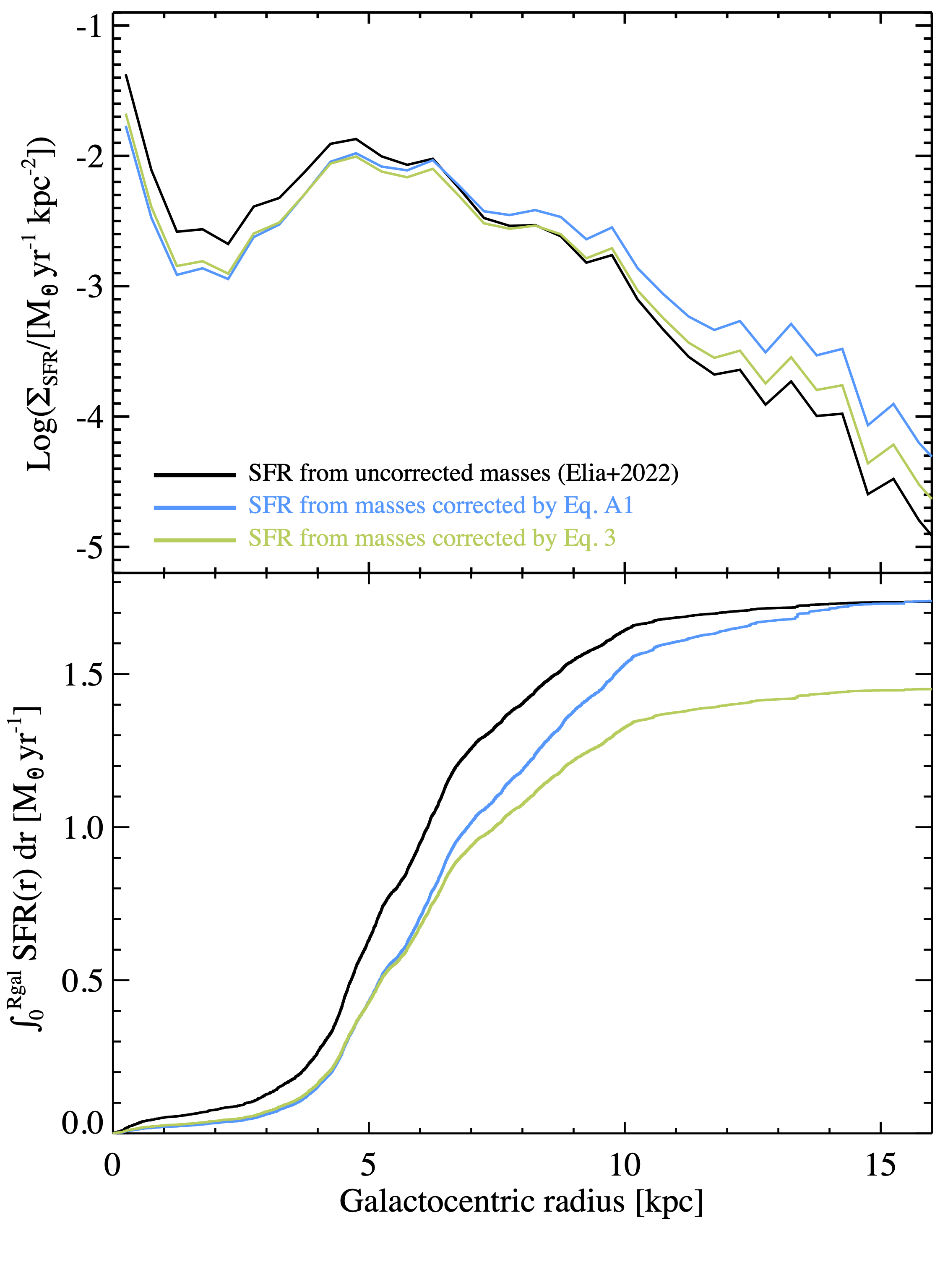}
\caption{\textit{Top}: Galactocentric radial profile of SFR density (in bins of 0.5~kpc). The black curve represents the SFR density $\Sigma_\mathrm{SFR}$ as calculated from clump masses through Eq.~\ref{sfrclump} \citep[already shown in][]{2022ApJ...941..162E}, while further corrections accounting for a variable gas-to-dust ratio suggested by Eqs.~\ref{gianneq} and~\ref{gtdelia} are represented with a blue and a green curve, respectively. The green curve is also represented in Fig.~\ref{sfr70profile}, where it is possible to appreciate typical error bars associated to values. Uncertainties on the other two curves are similar, so error bars are omitted for clarity. \textit{Bottom}: cumulative curves of profiles displayed in the top panel (same color coding).}
\label{corrprofile}
 \end{center}
\end{figure}

The top panel of Figure~\ref{corrprofile} shows, on the one hand, how the radial profile of $\Sigma_\mathrm{SFR}$ obtained from uncorrected Eq.~\ref{sfrclump} differs from those obtained by applying the correction in Eqs.~\ref{gianneq} and~\ref{gtdelia}: the first profile predicts higher star formation rates in the inner Galaxy, while the other two predict higher rates in the outer Galaxy. In particular, the profile corresponding to Eq.~\ref{gianneq} is the most distant from the uncorrected one, although the strong differences both in the inner and in the outer Galaxy get compensated in the global estimates of the SFR that they produce. 

On the other hand, the radial profiles given by Eqs.~\ref{sfrclump} and ~\ref{gtdelia} are more similar in the outer Galaxy, which, given the imbalance between the two in the inner Galaxy, finally determines the lower value of the $\mathrm{SFR}_\mathrm{H}$. 

These behaviors can be also appreciated by looking at the corresponding cumulative curves shown in the bottom panel of Figure~\ref{corrprofile}: the estimates based on the corrections given in Eqs.~\ref{gianneq} and~\ref{gtdelia} are almost indistinguishable up to $\rgal \sim 6$~kpc, after which the former steepens and finally ($\rgal \gtrsim 15$~kpc) reaches the cumulative value given by the uncorrected method.

\section{Checking for Systematic Issues}\label{appendix1}

In this appendix, we consider possible calibration errors and other systematic explanations for the severe underestimate of the SFR using the catalog values of the 70~\micron\ flux.
In general, Hi-GAL maps were cross-calibrated just with IRAS and PLANCK \citep{2010A&A...518L..88B}; however we wanted to make sure that there is no numerical glitch causing the systematic discrepancy that we noticed. To do that, we made a test by considering three coordinate square boxes in different Galactic locations, with sizes ranging from $1.7^\circ$ to $2^\circ$, and integrated the emission within them in the corresponding Hi-GAL map at 70~\micron, and IRAS maps at 60 and 100~\micron, respectively. Integrals were converted into Jy by taking into account different map pixel sizes. Locations and sizes of the boxes, together with total calculated fluxes at the three wavelengths reported in Table~\ref{irashigalcheck}. It can be seen that the total flux at 70~\micron\ is always comparable, within a factor of two, with the IRAS 60 and 100~\micron\ fluxes, and lies between these two. This clearly demonstrates that there is no wrong scaling affecting photometry of Hi-GAL 70~\micron\ maps.

\begin{table}
\begin{center}
\caption{Test fields (center coordinates and size) used to check possible incongruity between Hi-GAL 70~\micron\ map calibration and IRAS~60 and 100~\micron\ maps, and corresponding integrated flux at 60, 70, and 100~\micron, respectively.\label{irashigalcheck}}
\begin{tabular}{lcccccc}
\hline
& $\ell_c$ & $b_c$ & Size & $F_{60}$ & $F_{70}$ & $F_{100}$\\
& (\degree) & (\degree) & (\degree) & ($10^5$ Jy) & ($10^5$ Jy) & ($10^5$ Jy)\\
\hline
\hline
Field 1 & 19.2 & 0.2 & 1.7 & 3.7 & 5.2 & 12.9 \\
Field 2 & 70.3 & 1.1 & 1.9 & 0.8 & 1.3 &  2.5 \\
Field 3 & 330.0 & -0.2 & 2.0 & 4.2 & 5.7 & 15.8 \\
\hline
\end{tabular}
\end{center}
\end{table}

The next step was to compare source by source within the Hi-GAL catalog, to check, for example, whether the two estimates of the star formation rate agree when the star formation rate is high, or the 70~\micron\ method severely underestimates it when the star formation rate is low. This effect was seen for the 24~\micron\ method by
\citet{2013ApJ...765..129V}, who found that both the 24~\micron\ and total far-infrared methods underestimate the star formation rate in nearby clouds with low SFRs by a factor of 480 (median ratio).

\begin{figure}[ht!]
\includegraphics[width=0.5\textwidth]{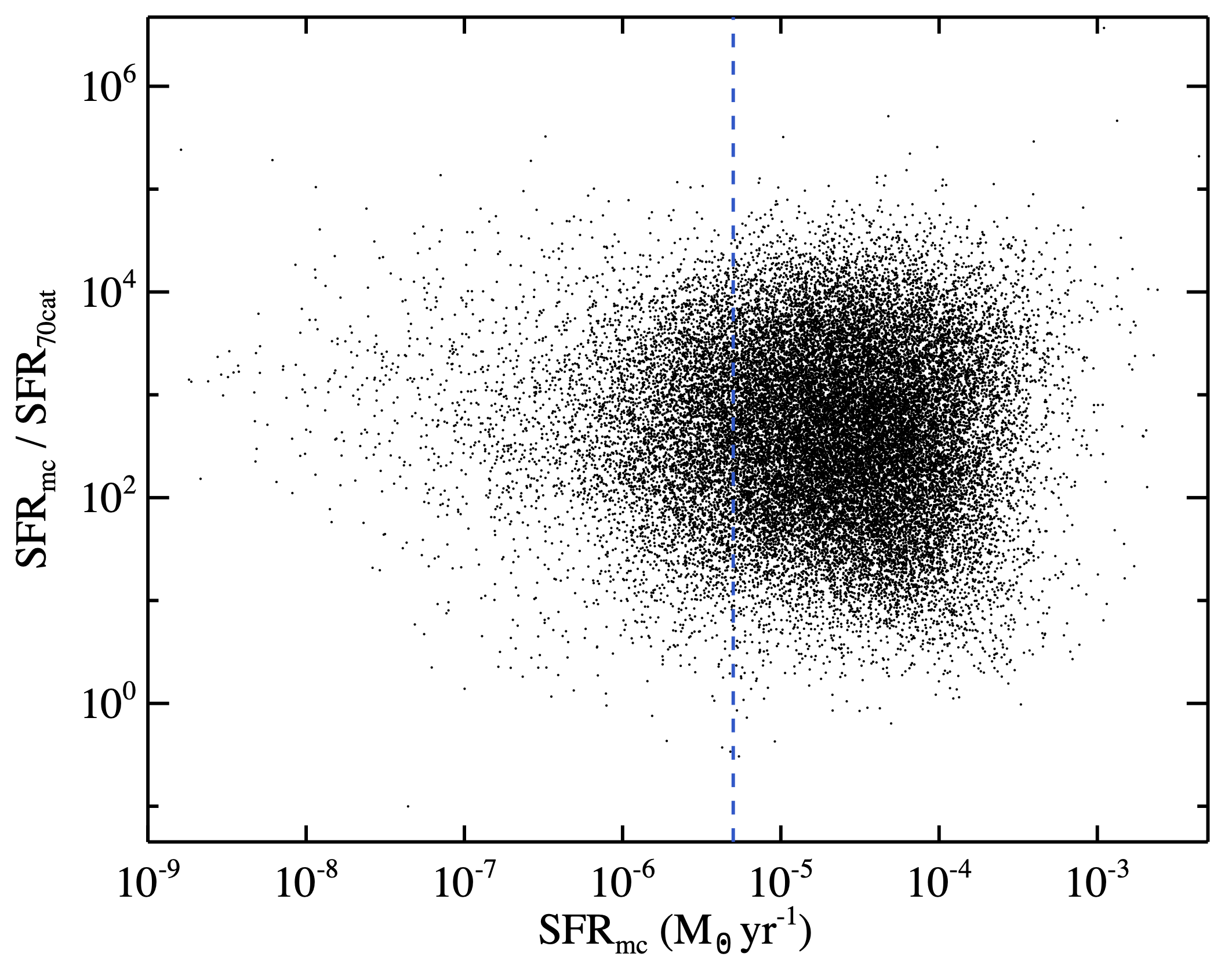}
\caption{The ratio of the SFR predicted from the molecular cloud model (\sfrmc) to the SFR predicted by 70~\micron\ (SFR$_{70\mathrm{cat}}$) is plotted versus \sfrmc. The blue dashed line indicates $\sfrmc > $5\ee{-6}$~\msunyr$ (see text).}
\label{l70vsmc}
\end{figure}

Considering Hi-GAL clumps classified as star forming and provided with a distance estimate, there are \nsourcesused\ sources to compare. The star formation rate (SFR$_{70\mathrm{cat}}$) based on the 70~\micron\ flux (from the catalog) was computed from Eq.~\ref{sfr70}, while the one based on clump mass \sfrmc\ was computed from Eq.~\ref{sfrclump}.

The ratio \sfrmc/SFR$_{70\mathrm{cat}}$ is plotted versus \sfrmc\ in Figure~\ref{l70vsmc}. The median of the ratio is \methodratiomedian. The large scatter of points in the plot does not suggest the presence of any trend.
\citet{2013ApJ...765..129V} 
found a very good correlation between radio continuum and total infrared estimates for SFRs greater than~$5\ee{-6}$~\msunyr.
If we consider the corresponding portion of the plot in Figure~\ref{l70vsmc} ($\sfrmc > 5\ee{-6}$~\msunyr), which corresponds to  \percentagehighsfr\% of points, no clear trend is seen, and a median ratio of \methodratiomedianhigh\ is found. 


\section{New Catalog of Molecular Clouds}\label{newcat}

A machine readable table is supplied with this paper, with columns described in Table \ref{tbl:newcat}. It reproduces the original catalog from
\citet{MD17}
but with distances assigned in this paper, as described in \S \ref{compother}. These new distances also change the columns for mass, height above the plane, physical area, and size. The masses are then further adjusted to account for the metallicity dependence of the luminosity-to-mass conversion factor, \alphaco. We use equation 6 for \alphaco(Z) and use the gradient in $Z$ (relative to the solar circle) from the [O/H] gradient of
\citet{2022MNRAS.510.4436M}:
\begin{equation}
\log Z = -0.044(\rgal - 8.34)\,.
\end{equation}
We add to the original table of
\citet{MD17}
two columns for SFR at the near and far kinematic distance, computed from the mass, free-fall time, and the \epsff\ from equation 7, again following EKO. We follow the example of 
\citet{MD17}
in supplying values for both near and far kinematic distance, while the flag shows our choice between them. This convention allows others to make different choices for resolving the kinematic distance ambiguity.
To summarize, all the clouds identified by 
\citet{MD17}
are still in the catalog, and columns 1 to 17 are identical to those in their catalog, columns 18 to 30 are modified by the new distance assignment, columns 27 to 30 are affected also by \alphaco($Z$), and columns 29 to 30 are newly added.

Several precautions should be considered in using this table. Some of the clouds in the original catalog with very low surface density are unlikely to be real 
\citep{2020ApJ...898....3L}. In addition, some distance assignments for clouds with anomalous velocities can be wrong, and this problem may erroneously place some clouds at very large \rgal. There are 29 clouds assigned an \rgal\ beyond 30~kpc, many in the direction of Galactic longitude between $-143\degree$ and $-153\degree$, most with low surface density ($< 6$~\msunpc) but very large size, and consequentially, mass, that are dubious.
Most do not contribute much to the SFR, but one in particular, catalog number 7787, claims a SFR of 1.69~\msunyr, larger than all the rest of the clouds combined! This cloud was assigned a distance of 88~kpc by \citet{MD17} and reassigned to a distance of 50~kpc by us. Its nominal mass of 6.7\ee7~\msun\ would be extreme for the far outer Galaxy. For reasons like this, we limit the summation over masses to 30~kpc, as noted in the main text. For continuity, we do not remove any clouds from the table, but reasonable cuts in \rgal\ will remove the dubious clouds from consideration. We follow \citet{MD17} in the format of table entries; we have no independent information on the precision with which the entries are known.

\begin{table}
  \begin{center}
\caption{\label{tab:catalog} Entries of the Molecular Cloud Catalog.}\label{tbl:newcat}
\tabskip=0pt
    \begin{tabular}{lll}\specialrule{\lightrulewidth}{0pt}{0pt} \specialrule{\lightrulewidth}{1.5pt}{\belowrulesep}
    Entry & Units & Description\\ \hline
    $Cloud$ & ... & Cloud number\\
    $N_{\rm comp}$ & ... & Number of Gaussian components\\
    $N_{\rm pix}$ & ... & Number of pixels on the sky\\
    $A$ & deg$^{2}$ & Angular area\\
    $l$ & deg & Barycentric Galactic longitude\\
    $e_l$ & deg & Galactic longitude standard deviation\\
    $b$ & deg & Barycentric Galactic latitude\\
    $e_b$ & deg & Galactic latitude standard deviation\\
    $\theta$ & deg & Angle with respect to $b=0^\circ$\\
    $W_{\rm CO}$ & K\,km\,s$^{-1}$ & Integrated CO emission\\
    $N_{\rm H2}$ & cm$^{-2}$ & Average column density\\
    $\Sigma$ & $M_\odot$\,pc$^{-2}$ & Surface density\\ 
    $v_{\rm cent}$ & km\,s$^{-1}$ & Centroid velocity\\
    $\sigma_v$  & km\,s$^{-1}$ & Velocity standard deviation\\
    $R_{\rm max}$ & deg & Largest eigenvalue of the inertia matrix\\
    $R_{\rm min}$ & deg & Smallest eigenvalue of the inertia matrix\\
    $R_{\rm ang}$ & deg & Angular size\\
    $R_{\rm gal}$ & kpc & Galactocentric radius\\
    $I_{\rm NF}$ & ... & Near or far distance flag\\
    $D_n$ & kpc & Near kinematic distance w R17 distance\\
    $D_f$ & kpc & Far kinematic distance w R17 distance\\
    $z_n$ & kpc & Near Distance to Galactic midplane w R17 distance\\
    $z_f$ & kpc & Far Distance to Galactic midplane w R17 distance\\
    $S_n$ & pc$^{2}$ & Near derived physical area w R17 distance \\
    $S_f$ & pc$^{2}$ & Far derived physical area w R17 distance \\
    $R_n$ & pc & Near derived physical size w R17 distancee\\
    $R_f$ & pc & Far derived physical size w R17 distancee\\
    $M_n$ & $M_\odot$ & Near derived mass w R17 distance\tablenotemark{1} \\
    $M_f$ & $M_\odot$ & Far derived mass w R17 distance\tablenotemark{1} \\ 
    $SFR_n$ & $M_\odot$ & Near SFR from EKO w R17 distance\tablenotemark{2} \\
    $SFR_f$ & $M_\odot$ & Far SFR from EKO w R17 distance\tablenotemark{2} \\
\bottomrule[\lightrulewidth]
      \end{tabular}
\end{center}
{{\bf Notes.} For clouds located in the inner Galaxy, two values are given for $z$, $S$, $R$, $M$, and SFR corresponding to the near and far kinematic distances. The index $I_{\rm NF}$ indicates which distance is used in this paper.
\tablenotetext{1}{ Uses CO luminosity to mass conversion of $\alphaco = 4.5 Z^{-0.8}$ and $\log(Z) = -0.044(\rgal - R_0)$.}
\tablenotetext{2}{Based on SFR = $\epsff M/\tff$ with $\epsff = 0.30 exp(-2.018 \alpha_{vir}^{0.5}$).}
}
  \end{table}


\bibliographystyle{aasjournal}
\clearpage
\bibliography{citejoin}

\begin{thebibliography}{}
\expandafter\ifx\csname natexlab\endcsname\relax\def\natexlab#1{#1}\fi
\providecommand{\url}[1]{\href{#1}{#1}}
\providecommand{\dodoi}[1]{doi:~\href{http://doi.org/#1}{\nolinkurl{#1}}}
\providecommand{\doeprint}[1]{\href{http://ascl.net/#1}{\nolinkurl{http://ascl.net/#1}}}
\providecommand{\doarXiv}[1]{\href{https://arxiv.org/abs/#1}{\nolinkurl{https://arxiv.org/abs/#1}}}

\bibitem[{{Aniano} {et~al.}(2020{\natexlab{a}}){Aniano}, {Draine}, {Hunt},
  {Sandstrom}, {Calzetti}, {Kennicutt}, {Dale}, {Galametz}, {Gordon}, {Leroy},
  {Smith}, {Roussel}, {Sauvage}, {Walter}, {Armus}, {Bolatto}, {Boquien},
  {Crocker}, {De Looze}, {Donovan Meyer}, {Helou}, {Hinz}, {Johnson}, {Koda},
  {Miller}, {Montiel}, {Murphy}, {Rela{\~n}o}, {Rix}, {Schinnerer}, {Skibba},
  {Wolfire}, \& {Engelbracht}}]{2020ApJ...889..150A}
{Aniano}, G., {Draine}, B.~T., {Hunt}, L.~K., {et~al.} 2020{\natexlab{a}},
  \apj, 889, 150, \dodoi{10.3847/1538-4357/ab5fdb}

\bibitem[{{Aniano} {et~al.}(2020{\natexlab{b}}){Aniano}, {Draine}, {Hunt},
  {Sandstrom}, {Calzetti}, {Kennicutt}, {Dale}, {Galametz}, {Gordon}, {Leroy},
  {Smith}, {Roussel}, {Sauvage}, {Walter}, {Armus}, {Bolatto}, {Boquien},
  {Crocker}, {De Looze}, {DDonovan Meyer}, {Helou}, {Hinz}, {Johnson}, {Koda},
  {Miller}, {Montiel}, {Murphy}, {Rela{\~n}o}, {Rix}, {Schinnerer}, {Skibba},
  {Wolfire}, \& {Engelbracht}}]{2020ApJ...897..184A}
---. 2020{\natexlab{b}}, \apj, 897, 184, \dodoi{10.3847/1538-4357/aba0bb}

\bibitem[{{Astropy Collaboration} {et~al.}(2013){Astropy Collaboration},
  {Robitaille}, {Tollerud}, {Greenfield}, {Droettboom}, {Bray}, {Aldcroft},
  {Davis}, {Ginsburg}, {Price-Whelan}, {Kerzendorf}, {Conley}, {Crighton},
  {Barbary}, {Muna}, {Ferguson}, {Grollier}, {Parikh}, {Nair}, {Unther},
  {Deil}, {Woillez}, {Conseil}, {Kramer}, {Turner}, {Singer}, {Fox}, {Weaver},
  {Zabalza}, {Edwards}, {Azalee Bostroem}, {Burke}, {Casey}, {Crawford},
  {Dencheva}, {Ely}, {Jenness}, {Labrie}, {Lim}, {Pierfederici}, {Pontzen},
  {Ptak}, {Refsdal}, {Servillat}, \& {Streicher}}]{2013A&A...558A..33A}
{Astropy Collaboration}, {Robitaille}, T.~P., {Tollerud}, E.~J., {et~al.} 2013,
  \aap, 558, A33, \dodoi{10.1051/0004-6361/201322068}

\bibitem[{{Astropy Collaboration} {et~al.}(2018){Astropy Collaboration},
  {Price-Whelan}, {Sip{\H{o}}cz}, {G{\"u}nther}, {Lim}, {Crawford}, {Conseil},
  {Shupe}, {Craig}, {Dencheva}, {Ginsburg}, {VanderPlas}, {Bradley},
  {P{\'e}rez-Su{\'a}rez}, {de Val-Borro}, {Aldcroft}, {Cruz}, {Robitaille},
  {Tollerud}, {Ardelean}, {Babej}, {Bach}, {Bachetti}, {Bakanov}, {Bamford},
  {Barentsen}, {Barmby}, {Baumbach}, {Berry}, {Biscani}, {Boquien}, {Bostroem},
  {Bouma}, {Brammer}, {Bray}, {Breytenbach}, {Buddelmeijer}, {Burke},
  {Calderone}, {Cano Rodr{\'\i}guez}, {Cara}, {Cardoso}, {Cheedella}, {Copin},
  {Corrales}, {Crichton}, {D'Avella}, {Deil}, {Depagne}, {Dietrich}, {Donath},
  {Droettboom}, {Earl}, {Erben}, {Fabbro}, {Ferreira}, {Finethy}, {Fox},
  {Garrison}, {Gibbons}, {Goldstein}, {Gommers}, {Greco}, {Greenfield},
  {Groener}, {Grollier}, {Hagen}, {Hirst}, {Homeier}, {Horton}, {Hosseinzadeh},
  {Hu}, {Hunkeler}, {Ivezi{\'c}}, {Jain}, {Jenness}, {Kanarek}, {Kendrew},
  {Kern}, {Kerzendorf}, {Khvalko}, {King}, {Kirkby}, {Kulkarni}, {Kumar},
  {Lee}, {Lenz}, {Littlefair}, {Ma}, {Macleod}, {Mastropietro}, {McCully},
  {Montagnac}, {Morris}, {Mueller}, {Mumford}, {Muna}, {Murphy}, {Nelson},
  {Nguyen}, {Ninan}, {N{\"o}the}, {Ogaz}, {Oh}, {Parejko}, {Parley}, {Pascual},
  {Patil}, {Patil}, {Plunkett}, {Prochaska}, {Rastogi}, {Reddy Janga},
  {Sabater}, {Sakurikar}, {Seifert}, {Sherbert}, {Sherwood-Taylor}, {Shih},
  {Sick}, {Silbiger}, {Singanamalla}, {Singer}, {Sladen}, {Sooley},
  {Sornarajah}, {Streicher}, {Teuben}, {Thomas}, {Tremblay}, {Turner},
  {Terr{\'o}n}, {van Kerkwijk}, {de la Vega}, {Watkins}, {Weaver}, {Whitmore},
  {Woillez}, {Zabalza}, \& {Astropy Contributors}}]{2018AJ....156..123A}
{Astropy Collaboration}, {Price-Whelan}, A.~M., {Sip{\H{o}}cz}, B.~M., {et~al.}
  2018, \aj, 156, 123, \dodoi{10.3847/1538-3881/aabc4f}

\bibitem[{{Astropy Collaboration} {et~al.}(2022){Astropy Collaboration},
  {Price-Whelan}, {Lim}, {Earl}, {Starkman}, {Bradley}, {Shupe}, {Patil},
  {Corrales}, {Brasseur}, {N{\"o}the}, {Donath}, {Tollerud}, {Morris},
  {Ginsburg}, {Vaher}, {Weaver}, {Tocknell}, {Jamieson}, {van Kerkwijk},
  {Robitaille}, {Merry}, {Bachetti}, {G{\"u}nther}, {Aldcroft},
  {Alvarado-Montes}, {Archibald}, {B{\'o}di}, {Bapat}, {Barentsen},
  {Baz{\'a}n}, {Biswas}, {Boquien}, {Burke}, {Cara}, {Cara}, {Conroy},
  {Conseil}, {Craig}, {Cross}, {Cruz}, {D'Eugenio}, {Dencheva}, {Devillepoix},
  {Dietrich}, {Eigenbrot}, {Erben}, {Ferreira}, {Foreman-Mackey}, {Fox},
  {Freij}, {Garg}, {Geda}, {Glattly}, {Gondhalekar}, {Gordon}, {Grant},
  {Greenfield}, {Groener}, {Guest}, {Gurovich}, {Handberg}, {Hart},
  {Hatfield-Dodds}, {Homeier}, {Hosseinzadeh}, {Jenness}, {Jones}, {Joseph},
  {Kalmbach}, {Karamehmetoglu}, {Ka{\l}uszy{\'n}ski}, {Kelley}, {Kern},
  {Kerzendorf}, {Koch}, {Kulumani}, {Lee}, {Ly}, {Ma}, {MacBride}, {Maljaars},
  {Muna}, {Murphy}, {Norman}, {O'Steen}, {Oman}, {Pacifici}, {Pascual},
  {Pascual-Granado}, {Patil}, {Perren}, {Pickering}, {Rastogi}, {Roulston},
  {Ryan}, {Rykoff}, {Sabater}, {Sakurikar}, {Salgado}, {Sanghi}, {Saunders},
  {Savchenko}, {Schwardt}, {Seifert-Eckert}, {Shih}, {Jain}, {Shukla}, {Sick},
  {Simpson}, {Singanamalla}, {Singer}, {Singhal}, {Sinha}, {Sip{\H{o}}cz},
  {Spitler}, {Stansby}, {Streicher}, {{\v{S}}umak}, {Swinbank}, {Taranu},
  {Tewary}, {Tremblay}, {de Val-Borro}, {Van Kooten}, {Vasovi{\'c}}, {Verma},
  {de Miranda Cardoso}, {Williams}, {Wilson}, {Winkel}, {Wood-Vasey}, {Xue},
  {Yoachim}, {Zhang}, {Zonca}, \& {Astropy Project
  Contributors}}]{2022ApJ...935..167A}
{Astropy Collaboration}, {Price-Whelan}, A.~M., {Lim}, P.~L., {et~al.} 2022,
  \apj, 935, 167, \dodoi{10.3847/1538-4357/ac7c74}

\bibitem[{{Barnes} {et~al.}(2017){Barnes}, {Longmore}, {Battersby}, {Bally},
  {Kruijssen}, {Henshaw}, \& {Walker}}]{2017MNRAS.469.2263B}
{Barnes}, A.~T., {Longmore}, S.~N., {Battersby}, C., {et~al.} 2017, \mnras,
  469, 2263, \dodoi{10.1093/mnras/stx941}

\bibitem[{{Battersby} {et~al.}(2020){Battersby}, {Keto}, {Walker}, {Barnes},
  {Callanan}, {Ginsburg}, {Hatchfield}, {Henshaw}, {Kauffmann}, {Kruijssen},
  {Longmore}, {Lu}, {Mills}, {Pillai}, {Zhang}, {Bally}, {Butterfield},
  {Contreras}, {Ho}, {Ott}, {Patel}, \& {Tolls}}]{2020ApJS..249...35B}
{Battersby}, C., {Keto}, E., {Walker}, D., {et~al.} 2020, \apjs, 249, 35,
  \dodoi{10.3847/1538-4365/aba18e}

\bibitem[{{Benjamin} {et~al.}(2005){Benjamin}, {Churchwell}, {Babler},
  {Indebetouw}, {Meade}, {Whitney}, {Watson}, {Wolfire}, {Wolff}, {Ignace},
  {Bania}, {Bracker}, {Clemens}, {Chomiuk}, {Cohen}, {Dickey}, {Jackson},
  {Kobulnicky}, {Mercer}, {Mathis}, {Stolovy}, \&
  {Uzpen}}]{2005ApJ...630L.149B}
{Benjamin}, R.~A., {Churchwell}, E., {Babler}, B.~L., {et~al.} 2005, \apjl,
  630, L149, \dodoi{10.1086/491785}

\bibitem[{{Bergin} \& {Tafalla}(2007)}]{2007ARA&A..45..339B}
{Bergin}, E.~A., \& {Tafalla}, M. 2007, \araa, 45, 339,
  \dodoi{10.1146/annurev.astro.45.071206.100404}

\bibitem[{{Bernard} {et~al.}(2010){Bernard}, {Paradis}, {Marshall}, {Montier},
  {Lagache}, {Paladini}, {Veneziani}, {Brunt}, {Mottram}, {Martin},
  {Ristorcelli}, {Noriega-Crespo}, {Compi{\`e}gne}, {Flagey}, {Anderson},
  {Popescu}, {Tuffs}, {Reach}, {White}, {Benedettini}, {Calzoletti},
  {Digiorgio}, {Faustini}, {Juvela}, {Joblin}, {Joncas}, {Mivilles-Deschenes},
  {Olmi}, {Traficante}, {Piacentini}, {Zavagno}, \&
  {Molinari}}]{2010A&A...518L..88B}
{Bernard}, J.~P., {Paradis}, D., {Marshall}, D.~J., {et~al.} 2010, \aap, 518,
  L88, \dodoi{10.1051/0004-6361/201014540}

\bibitem[{{Bigiel} {et~al.}(2011){Bigiel}, {Leroy}, {Walter}, {Brinks}, {de
  Blok}, {Kramer}, {Rix}, {Schruba}, {Schuster}, {Usero}, \&
  {Wiesemeyer}}]{2011ApJ...730L..13B}
{Bigiel}, F., {Leroy}, A.~K., {Walter}, F., {et~al.} 2011, \apjl, 730, L13,
  \dodoi{10.1088/2041-8205/730/2/L13}

\bibitem[{{Bland-Hawthorn} \& {Gerhard}(2016)}]{2016ARA&A..54..529B}
{Bland-Hawthorn}, J., \& {Gerhard}, O. 2016, \araa, 54, 529,
  \dodoi{10.1146/annurev-astro-081915-023441}

\bibitem[{{Calzetti} {et~al.}(2007){Calzetti}, {Kennicutt}, {Engelbracht},
  {Leitherer}, {Draine}, {Kewley}, {Moustakas}, {Sosey}, {Dale}, {Gordon},
  {Helou}, {Hollenbach}, {Armus}, {Bendo}, {Bot}, {Buckalew}, {Jarrett}, {Li},
  {Meyer}, {Murphy}, {Prescott}, {Regan}, {Rieke}, {Roussel}, {Sheth}, {Smith},
  {Thornley}, \& {Walter}}]{2007ApJ...666..870C}
{Calzetti}, D., {Kennicutt}, R.~C., {Engelbracht}, C.~W., {et~al.} 2007, \apj,
  666, 870, \dodoi{10.1086/520082}

\bibitem[{{Calzetti} {et~al.}(2010){Calzetti}, {Wu}, {Hong}, {Kennicutt},
  {Lee}, {Dale}, {Engelbracht}, {van Zee}, {Draine}, {Hao}, {Gordon},
  {Moustakas}, {Murphy}, {Regan}, {Begum}, {Block}, {Dalcanton}, {Funes}, {Gil
  de Paz}, {Johnson}, {Sakai}, {Skillman}, {Walter}, {Weisz}, {Williams}, \&
  {Wu}}]{2010ApJ...714.1256C}
{Calzetti}, D., {Wu}, S.~Y., {Hong}, S., {et~al.} 2010, \apj, 714, 1256,
  \dodoi{10.1088/0004-637X/714/2/1256}

\bibitem[{{Chomiuk} \& {Povich}(2011)}]{2011AJ....142..197C}
{Chomiuk}, L., \& {Povich}, M.~S. 2011, \aj, 142, 197,
  \dodoi{10.1088/0004-6256/142/6/197}

\bibitem[{{Crocker} {et~al.}(2011){Crocker}, {Jones}, {Aharonian}, {Law},
  {Melia}, {Oka}, \& {Ott}}]{2011MNRAS.413..763C}
{Crocker}, R.~M., {Jones}, D.~I., {Aharonian}, F., {et~al.} 2011, \mnras, 413,
  763, \dodoi{10.1111/j.1365-2966.2010.18170.x}

\bibitem[{{da Silva} {et~al.}(2014){da Silva}, {Krumholz}, {Fumagalli}, \&
  {Fall}}]{2014MNRAS.438.2355D}
{da Silva}, R.~L., {Krumholz}, M.~R., {Fumagalli}, M., \& {Fall}, S.~M. 2014,
  \mnras, 438, 2355, \dodoi{10.1093/mnras/stt2351}

\bibitem[{{Dame} {et~al.}(2001){Dame}, {Hartmann}, \& {Thaddeus}}]{Dame01}
{Dame}, T.~M., {Hartmann}, D., \& {Thaddeus}, P. 2001, \apj, 547, 792,
  \dodoi{10.1086/318388}

\bibitem[{{Draine} {et~al.}(2007){Draine}, {Dale}, {Bendo}, {Gordon}, {Smith},
  {Armus}, {Engelbracht}, {Helou}, {Kennicutt}, {Li}, {Roussel}, {Walter},
  {Calzetti}, {Moustakas}, {Murphy}, {Rieke}, {Bot}, {Hollenbach}, {Sheth}, \&
  {Teplitz}}]{2007ApJ...663..866D}
{Draine}, B.~T., {Dale}, D.~A., {Bendo}, G., {et~al.} 2007, \apj, 663, 866,
  \dodoi{10.1086/518306}

\bibitem[{{Elia} \& {Pezzuto}(2016)}]{eli16}
{Elia}, D., \& {Pezzuto}, S. 2016, \mnras, 461, 1328,
  \dodoi{10.1093/mnras/stw1399}

\bibitem[{{Elia} {et~al.}(2017){Elia}, {Molinari}, {Schisano}, {Pestalozzi},
  {Pezzuto}, {Merello}, {Noriega-Crespo}, {Moore}, {Russeil}, {Mottram},
  {Paladini}, {Strafella}, {Benedettini}, {Bernard}, {Di Giorgio}, {Eden},
  {Fukui}, {Plume}, {Bally}, {Martin}, {Ragan}, {Jaffa}, {Motte}, {Olmi},
  {Schneider}, {Testi}, {Wyrowski}, {Zavagno}, {Calzoletti}, {Faustini},
  {Natoli}, {Palmeirim}, {Piacentini}, {Piazzo}, {Pilbratt}, {Polychroni},
  {Baldeschi}, {Beltr{\'a}n}, {Billot}, {Cambr{\'e}sy}, {Cesaroni},
  {Garc{\'\i}a-Lario}, {Hoare}, {Huang}, {Joncas}, {Liu}, {Maiolo}, {Marsh},
  {Maruccia}, {M{\`e}ge}, {Peretto}, {Rygl}, {Schilke}, {Thompson},
  {Traficante}, {Umana}, {Veneziani}, {Ward-Thompson}, {Whitworth}, {Arab},
  {Bandieramonte}, {Becciani}, {Brescia}, {Buemi}, {Bufano}, {Butora},
  {Cavuoti}, {Costa}, {Fiorellino}, {Hajnal}, {Hayakawa}, {Kacsuk}, {Leto}, {Li
  Causi}, {Marchili}, {Martinavarro-Armengol}, {Mercurio}, {Molinaro},
  {Riccio}, {Sano}, {Sciacca}, {Tachihara}, {Torii}, {Trigilio}, {Vitello}, \&
  {Yamamoto}}]{2017MNRAS.471..100E}
{Elia}, D., {Molinari}, S., {Schisano}, E., {et~al.} 2017, \mnras, 471, 100,
  \dodoi{10.1093/mnras/stx1357}

\bibitem[{{Elia} {et~al.}(2021){Elia}, {Merello}, {Molinari}, {Schisano},
  {Zavagno}, {Russeil}, {M{\`e}ge}, {Martin}, {Olmi}, {Pestalozzi}, {Plume},
  {Ragan}, {Benedettini}, {Eden}, {Moore}, {Noriega-Crespo}, {Paladini},
  {Palmeirim}, {Pezzuto}, {Pilbratt}, {Rygl}, {Schilke}, {Strafella}, {Tan},
  {Traficante}, {Baldeschi}, {Bally}, {di Giorgio}, {Fiorellino}, {Liu},
  {Piazzo}, \& {Polychroni}}]{2021MNRAS.504.2742E}
{Elia}, D., {Merello}, M., {Molinari}, S., {et~al.} 2021, \mnras, 504, 2742,
  \dodoi{10.1093/mnras/stab1038}

\bibitem[{{Elia} {et~al.}(2022){Elia}, {Molinari}, {Schisano}, {Soler},
  {Merello}, {Russeil}, {Veneziani}, {Zavagno}, {Noriega-Crespo}, {Olmi},
  {Benedettini}, {Hennebelle}, {Klessen}, {Leurini}, {Paladini}, {Pezzuto},
  {Traficante}, {Eden}, {Martin}, {Sormani}, {Coletta}, {Colman}, {Plume},
  {Maruccia}, {Mininni}, \& {Liu}}]{2022ApJ...941..162E}
{Elia}, D., {Molinari}, S., {Schisano}, E., {et~al.} 2022, \apj, 941, 162,
  \dodoi{10.3847/1538-4357/aca27d}

\bibitem[{{Evans} {et~al.}(2022){Evans}, {Kim}, \&
  {Ostriker}}]{2022ApJ...929L..18E}
{Evans}, N.~J., {Kim}, J.-G., \& {Ostriker}, E.~C. 2022, \apjl, 929, L18,
  \dodoi{10.3847/2041-8213/ac6427}

\bibitem[{{Evans} {et~al.}(2020){Evans}, {Kim}, {Wu}, {Chao}, {Heyer}, {Liu},
  {Nguyen-Lu'o'ng}, \& {Kauffmann}}]{2020ApJ...894..103E}
{Evans}, II, N.~J., {Kim}, K.-T., {Wu}, J., {et~al.} 2020, \apj, 894, 103,
  \dodoi{10.3847/1538-4357/ab8938}

\bibitem[{{Gao} \& {Solomon}(2004)}]{2004ApJ...606..271G}
{Gao}, Y., \& {Solomon}, P.~M. 2004, \apj, 606, 271, \dodoi{10.1086/382999}

\bibitem[{{Giannetti} {et~al.}(2017){Giannetti}, {Leurini}, {K{\"o}nig},
  {Urquhart}, {Pillai}, {Brand}, {Kauffmann}, {Wyrowski}, \&
  {Menten}}]{2017A&A...606L..12G}
{Giannetti}, A., {Leurini}, S., {K{\"o}nig}, C., {et~al.} 2017, \aap, 606, L12,
  \dodoi{10.1051/0004-6361/201731728}

\bibitem[{{Griffin} {et~al.}(2010){Griffin}, {Abergel}, {Abreu}, {Ade},
  {Andr{\'e}}, {Augueres}, {Babbedge}, {Bae}, {Baillie}, {Baluteau}, {Barlow},
  {Bendo}, {Benielli}, {Bock}, {Bonhomme}, {Brisbin}, {Brockley-Blatt},
  {Caldwell}, {Cara}, {Castro-Rodriguez}, {Cerulli}, {Chanial}, {Chen},
  {Clark}, {Clements}, {Clerc}, {Coker}, {Communal}, {Conversi}, {Cox},
  {Crumb}, {Cunningham}, {Daly}, {Davis}, {de Antoni}, {Delderfield}, {Devin},
  {di Giorgio}, {Didschuns}, {Dohlen}, {Donati}, {Dowell}, {Dowell}, {Duband},
  {Dumaye}, {Emery}, {Ferlet}, {Ferrand}, {Fontignie}, {Fox}, {Franceschini},
  {Frerking}, {Fulton}, {Garcia}, {Gastaud}, {Gear}, {Glenn}, {Goizel},
  {Griffin}, {Grundy}, {Guest}, {Guillemet}, {Hargrave}, {Harwit}, {Hastings},
  {Hatziminaoglou}, {Herman}, {Hinde}, {Hristov}, {Huang}, {Imhof}, {Isaak},
  {Israelsson}, {Ivison}, {Jennings}, {Kiernan}, {King}, {Lange}, {Latter},
  {Laurent}, {Laurent}, {Leeks}, {Lellouch}, {Levenson}, {Li}, {Li},
  {Lilienthal}, {Lim}, {Liu}, {Lu}, {Madden}, {Mainetti}, {Marliani}, {McKay},
  {Mercier}, {Molinari}, {Morris}, {Moseley}, {Mulder}, {Mur}, {Naylor},
  {Nguyen}, {O'Halloran}, {Oliver}, {Olofsson}, {Olofsson}, {Orfei}, {Page},
  {Pain}, {Panuzzo}, {Papageorgiou}, {Parks}, {Parr-Burman}, {Pearce},
  {Pearson}, {P{\'e}rez-Fournon}, {Pinsard}, {Pisano}, {Podosek}, {Pohlen},
  {Polehampton}, {Pouliquen}, {Rigopoulou}, {Rizzo}, {Roseboom}, {Roussel},
  {Rowan-Robinson}, {Rownd}, {Saraceno}, {Sauvage}, {Savage}, {Savini},
  {Sawyer}, {Scharmberg}, {Schmitt}, {Schneider}, {Schulz}, {Schwartz},
  {Shafer}, {Shupe}, {Sibthorpe}, {Sidher}, {Smith}, {Smith}, {Smith},
  {Spencer}, {Stobie}, {Sudiwala}, {Sukhatme}, {Surace}, {Stevens}, {Swinyard},
  {Trichas}, {Tourette}, {Triou}, {Tseng}, {Tucker}, {Turner}, {Vaccari},
  {Valtchanov}, {Vigroux}, {Virique}, {Voellmer}, {Walker}, {Ward}, {Waskett},
  {Weilert}, {Wesson}, {White}, {Whitehouse}, {Wilson}, {Winter}, {Woodcraft},
  {Wright}, {Xu}, {Zavagno}, {Zemcov}, {Zhang}, \&
  {Zonca}}]{2010A&A...518L...3G}
{Griffin}, M.~J., {Abergel}, A., {Abreu}, A., {et~al.} 2010, \aap, 518, L3,
  \dodoi{10.1051/0004-6361/201014519}

\bibitem[{{Grudi{\'c}} {et~al.}(2022){Grudi{\'c}}, {Guszejnov}, {Offner},
  {Rosen}, {Raju}, {Faucher-Gigu{\`e}re}, \& {Hopkins}}]{2022MNRAS.512..216G}
{Grudi{\'c}}, M.~Y., {Guszejnov}, D., {Offner}, S. S.~R., {et~al.} 2022,
  \mnras, 512, 216, \dodoi{10.1093/mnras/stac526}

\bibitem[{{Heiderman} {et~al.}(2010){Heiderman}, {Evans}, {Allen}, {Huard}, \&
  {Heyer}}]{2010ApJ...723.1019H}
{Heiderman}, A., {Evans}, II, N.~J., {Allen}, L.~E., {Huard}, T., \& {Heyer},
  M. 2010, \apj, 723, 1019, \dodoi{10.1088/0004-637X/723/2/1019}

\bibitem[{{Henshaw} {et~al.}(2023){Henshaw}, {Barnes}, {Battersby}, {Ginsburg},
  {Sormani}, \& {Walker}}]{2023ASPC..534...83H}
{Henshaw}, J.~D., {Barnes}, A.~T., {Battersby}, C., {et~al.} 2023, in
  Astronomical Society of the Pacific Conference Series, Vol. 534, Protostars
  and Planets VII, ed. S.~{Inutsuka}, Y.~{Aikawa}, T.~{Muto}, K.~{Tomida}, \&
  M.~{Tamura}, 83

\bibitem[{{Hu} {et~al.}(2024){Hu}, {Wibking}, {Krumholz}, \&
  {Federrath}}]{2024MNRAS.534.2426H}
{Hu}, Z., {Wibking}, B.~D., {Krumholz}, M.~R., \& {Federrath}, C. 2024, \mnras,
  534, 2426, \dodoi{10.1093/mnras/stae2241}

\bibitem[{{Immer} {et~al.}(2012){Immer}, {Schuller}, {Omont}, \&
  {Menten}}]{2012A&A...537A.121I}
{Immer}, K., {Schuller}, F., {Omont}, A., \& {Menten}, K.~M. 2012, \aap, 537,
  A121, \dodoi{10.1051/0004-6361/201117857}

\bibitem[{{Kennicutt}(1998)}]{1998ApJ...498..541K}
{Kennicutt}, Robert~C., J. 1998, \apj, 498, 541, \dodoi{10.1086/305588}

\bibitem[{{Kennicutt} {et~al.}(2003){Kennicutt}, {Armus}, {Bendo}, {Calzetti},
  {Dale}, {Draine}, {Engelbracht}, {Gordon}, {Grauer}, {Helou}, {Hollenbach},
  {Jarrett}, {Kewley}, {Leitherer}, {Li}, {Malhotra}, {Regan}, {Rieke},
  {Rieke}, {Roussel}, {Smith}, {Thornley}, \& {Walter}}]{2003PASP..115..928K}
{Kennicutt}, Robert~C., J., {Armus}, L., {Bendo}, G., {et~al.} 2003, \pasp,
  115, 928, \dodoi{10.1086/376941}

\bibitem[{{Kennicutt} \& {Evans}(2012)}]{2012ARA&A..50..531K}
{Kennicutt}, R.~C., \& {Evans}, N.~J. 2012, \araa, 50, 531,
  \dodoi{10.1146/annurev-astro-081811-125610}

\bibitem[{{Kim} {et~al.}(2023){Kim}, {Gong}, {Kim}, \&
  {Ostriker}}]{2023ApJS..264...10K}
{Kim}, J.-G., {Gong}, M., {Kim}, C.-G., \& {Ostriker}, E.~C. 2023, \apjs, 264,
  10, \dodoi{10.3847/1538-4365/ac9b1d}

\bibitem[{{Kruijssen} {et~al.}(2014){Kruijssen}, {Longmore}, {Elmegreen},
  {Murray}, {Bally}, {Testi}, \& {Kennicutt}}]{2014MNRAS.440.3370K}
{Kruijssen}, J.~M.~D., {Longmore}, S.~N., {Elmegreen}, B.~G., {et~al.} 2014,
  \mnras, 440, 3370, \dodoi{10.1093/mnras/stu494}

\bibitem[{{Lada} \& {Dame}(2020)}]{2020ApJ...898....3L}
{Lada}, C.~J., \& {Dame}, T.~M. 2020, \apj, 898, 3,
  \dodoi{10.3847/1538-4357/ab9bfb}

\bibitem[{{Lada} {et~al.}(2012){Lada}, {Forbrich}, {Lombardi}, \&
  {Alves}}]{2012ApJ...745..190L}
{Lada}, C.~J., {Forbrich}, J., {Lombardi}, M., \& {Alves}, J.~F. 2012, \apj,
  745, 190, \dodoi{10.1088/0004-637X/745/2/190}

\bibitem[{{Lada} {et~al.}(2010){Lada}, {Lombardi}, \&
  {Alves}}]{2010ApJ...724..687L}
{Lada}, C.~J., {Lombardi}, M., \& {Alves}, J.~F. 2010, \apj, 724, 687,
  \dodoi{10.1088/0004-637X/724/1/687}

\bibitem[{{Lawton} {et~al.}(2010){Lawton}, {Gordon}, {Babler}, {Block},
  {Bolatto}, {Bracker}, {Carlson}, {Engelbracht}, {Hora}, {Indebetouw},
  {Madden}, {Meade}, {Meixner}, {Misselt}, {Oey}, {Oliveira}, {Robitaille},
  {Sewilo}, {Shiao}, {Vijh}, \& {Whitney}}]{2010ApJ...716..453L}
{Lawton}, B., {Gordon}, K.~D., {Babler}, B., {et~al.} 2010, \apj, 716, 453,
  \dodoi{10.1088/0004-637X/716/1/453}

\bibitem[{{Lee} {et~al.}(2024){Lee}, {Koda}, {Hirota}, {Egusa}, \&
  {Heyer}}]{2024arXiv240414503L}
{Lee}, A.~M., {Koda}, J., {Hirota}, A., {Egusa}, F., \& {Heyer}, M. 2024, arXiv
  e-prints, arXiv:2404.14503, \dodoi{10.48550/arXiv.2404.14503}

\bibitem[{{Li} {et~al.}(2010){Li}, {Calzetti}, {Kennicutt}, {Hong},
  {Engelbracht}, {Dale}, \& {Moustakas}}]{2010ApJ...725..677L}
{Li}, Y., {Calzetti}, D., {Kennicutt}, R.~C., {et~al.} 2010, \apj, 725, 677,
  \dodoi{10.1088/0004-637X/725/1/677}

\bibitem[{{Lian} {et~al.}(2024){Lian}, {Zasowski}, {Chen}, {Imig}, {Wang},
  {Boardman}, \& {Liu}}]{2024NatAs.tmp..190L}
{Lian}, J., {Zasowski}, G., {Chen}, B., {et~al.} 2024, Nature Astronomy,
  \dodoi{10.1038/s41550-024-02315-7}

\bibitem[{{Licquia} \& {Newman}(2015)}]{2015ApJ...806...96L}
{Licquia}, T.~C., \& {Newman}, J.~A. 2015, \apj, 806, 96,
  \dodoi{10.1088/0004-637X/806/1/96}

\bibitem[{{Linzer} {et~al.}(2024){Linzer}, {Kim}, {Kim}, \&
  {Ostriker}}]{2024ApJ...975..173L}
{Linzer}, N.~B., {Kim}, J.-G., {Kim}, C.-G., \& {Ostriker}, E.~C. 2024, \apj,
  975, 173, \dodoi{10.3847/1538-4357/ad7733}

\bibitem[{{Longmore} {et~al.}(2013){Longmore}, {Bally}, {Testi}, {Purcell},
  {Walsh}, {Bressert}, {Pestalozzi}, {Molinari}, {Ott}, {Cortese}, {Battersby},
  {Murray}, {Lee}, {Kruijssen}, {Schisano}, \& {Elia}}]{2013MNRAS.429..987L}
{Longmore}, S.~N., {Bally}, J., {Testi}, L., {et~al.} 2013, \mnras, 429, 987,
  \dodoi{10.1093/mnras/sts376}

\bibitem[{{Madau} \& {Dickinson}(2014)}]{2014ARA&A..52..415M}
{Madau}, P., \& {Dickinson}, M. 2014, \araa, 52, 415,
  \dodoi{10.1146/annurev-astro-081811-125615}

\bibitem[{{McKee} \& {Ostriker}(2007)}]{2007ARA&A..45..565M}
{McKee}, C.~F., \& {Ostriker}, E.~C. 2007, \araa, 45, 565,
  \dodoi{10.1146/annurev.astro.45.051806.110602}

\bibitem[{{M{\`e}ge} {et~al.}(2021){M{\`e}ge}, {Russeil}, {Zavagno}, {Elia},
  {Molinari}, {Brunt}, {Butora}, {Cambresy}, {Di Giorgio}, {Fenouillet},
  {Fukui}, {Lambert}, {Makai}, {Merello}, {Meunier}, {Molinaro}, {Moreau},
  {Pezzuto}, {Poulin}, {Schisano}, \& {Schuller}}]{2021A&A...646A..74M}
{M{\`e}ge}, P., {Russeil}, D., {Zavagno}, A., {et~al.} 2021, \aap, 646, A74,
  \dodoi{10.1051/0004-6361/202038956}

\bibitem[{{M{\'e}ndez-Delgado} {et~al.}(2022){M{\'e}ndez-Delgado}, {Amayo},
  {Arellano-C{\'o}rdova}, {Esteban}, {Garc{\'\i}a-Rojas}, {Carigi}, \&
  {Delgado-Inglada}}]{2022MNRAS.510.4436M}
{M{\'e}ndez-Delgado}, J.~E., {Amayo}, A., {Arellano-C{\'o}rdova}, K.~Z.,
  {et~al.} 2022, \mnras, 510, 4436, \dodoi{10.1093/mnras/stab3782}

\bibitem[{{Miville-Desch{\^e}nes} {et~al.}(2017){Miville-Desch{\^e}nes},
  {Murray}, \& {Lee}}]{MD17}
{Miville-Desch{\^e}nes}, M.-A., {Murray}, N., \& {Lee}, E.~J. 2017, \apj, 834,
  57, \dodoi{10.3847/1538-4357/834/1/57}

\bibitem[{{Molinari} {et~al.}(2010){Molinari}, {Swinyard}, {Bally}, {Barlow},
  {Bernard}, {Martin}, {Moore}, {Noriega-Crespo}, {Plume}, {Testi}, {Zavagno},
  {Abergel}, {Ali}, {Andr{\'e}}, {Baluteau}, {Benedettini}, {Bern{\'e}},
  {Billot}, {Blommaert}, {Bontemps}, {Boulanger}, {Brand}, {Brunt}, {Burton},
  {Campeggio}, {Carey}, {Caselli}, {Cesaroni}, {Cernicharo}, {Chakrabarti},
  {Chrysostomou}, {Codella}, {Cohen}, {Compiegne}, {Davis}, {de Bernardis}, {de
  Gasperis}, {Di Francesco}, {di Giorgio}, {Elia}, {Faustini}, {Fischera},
  {Fukui}, {Fuller}, {Ganga}, {Garcia-Lario}, {Giard}, {Giardino}, {Glenn},
  {Goldsmith}, {Griffin}, {Hoare}, {Huang}, {Jiang}, {Joblin}, {Joncas},
  {Juvela}, {Kirk}, {Lagache}, {Li}, {Lim}, {Lord}, {Lucas}, {Maiolo},
  {Marengo}, {Marshall}, {Masi}, {Massi}, {Matsuura}, {Meny}, {Minier},
  {Miville-Desch{\^e}nes}, {Montier}, {Motte}, {M{\"u}ller}, {Natoli}, {Neves},
  {Olmi}, {Paladini}, {Paradis}, {Pestalozzi}, {Pezzuto}, {Piacentini},
  {Pomar{\`e}s}, {Popescu}, {Reach}, {Richer}, {Ristorcelli}, {Roy}, {Royer},
  {Russeil}, {Saraceno}, {Sauvage}, {Schilke}, {Schneider-Bontemps},
  {Schuller}, {Schultz}, {Shepherd}, {Sibthorpe}, {Smith}, {Smith},
  {Spinoglio}, {Stamatellos}, {Strafella}, {Stringfellow}, {Sturm}, {Taylor},
  {Thompson}, {Tuffs}, {Umana}, {Valenziano}, {Vavrek}, {Viti}, {Waelkens},
  {Ward-Thompson}, {White}, {Wyrowski}, {Yorke}, \&
  {Zhang}}]{2010PASP..122..314M}
{Molinari}, S., {Swinyard}, B., {Bally}, J., {et~al.} 2010, \pasp, 122, 314,
  \dodoi{10.1086/651314}

\bibitem[{{Molinari} {et~al.}(2016){Molinari}, {Schisano}, {Elia},
  {Pestalozzi}, {Traficante}, {Pezzuto}, {Swinyard}, {Noriega-Crespo}, {Bally},
  {Moore}, {Plume}, {Zavagno}, {di Giorgio}, {Liu}, {Pilbratt}, {Mottram},
  {Russeil}, {Piazzo}, {Veneziani}, {Benedettini}, {Calzoletti}, {Faustini},
  {Natoli}, {Piacentini}, {Merello}, {Palmese}, {Del Grande}, {Polychroni},
  {Rygl}, {Polenta}, {Barlow}, {Bernard}, {Martin}, {Testi}, {Ali},
  {Andr{\'e}}, {Beltr{\'a}n}, {Billot}, {Carey}, {Cesaroni}, {Compi{\`e}gne},
  {Eden}, {Fukui}, {Garcia-Lario}, {Hoare}, {Huang}, {Joncas}, {Lim}, {Lord},
  {Martinavarro-Armengol}, {Motte}, {Paladini}, {Paradis}, {Peretto},
  {Robitaille}, {Schilke}, {Schneider}, {Schulz}, {Sibthorpe}, {Strafella},
  {Thompson}, {Umana}, {Ward-Thompson}, \& {Wyrowski}}]{2016A&A...591A.149M}
{Molinari}, S., {Schisano}, E., {Elia}, D., {et~al.} 2016, \aap, 591, A149,
  \dodoi{10.1051/0004-6361/201526380}

\bibitem[{{Nakanishi} \& {Sofue}(2016)}]{2016PASJ...68....5N}
{Nakanishi}, H., \& {Sofue}, Y. 2016, \pasj, 68, 5, \dodoi{10.1093/pasj/psv108}

\bibitem[{{Oka} {et~al.}(2001){Oka}, {Hasegawa}, {Sato}, {Tsuboi}, {Miyazaki},
  \& {Sugimoto}}]{2001ApJ...562..348O}
{Oka}, T., {Hasegawa}, T., {Sato}, F., {et~al.} 2001, \apj, 562, 348,
  \dodoi{10.1086/322976}

\bibitem[{{Patra} {et~al.}(2022){Patra}, {Evans}, {Kim}, {Heyer}, {Kauffmann},
  {Jose}, {Samal}, \& {Das}}]{2022AJ....164..129P}
{Patra}, S., {Evans}, II, N.~J., {Kim}, K.-T., {et~al.} 2022, \aj, 164, 129,
  \dodoi{10.3847/1538-3881/ac83af}

\bibitem[{{Pilbratt} {et~al.}(2010){Pilbratt}, {Riedinger}, {Passvogel},
  {Crone}, {Doyle}, {Gageur}, {Heras}, {Jewell}, {Metcalfe}, {Ott}, \&
  {Schmidt}}]{2010A&A...518L...1P}
{Pilbratt}, G.~L., {Riedinger}, J.~R., {Passvogel}, T., {et~al.} 2010, \aap,
  518, L1, \dodoi{10.1051/0004-6361/201014759}

\bibitem[{{Pineda} {et~al.}(2024){Pineda}, {Horiuchi}, {Anderson}, {Luisi},
  {Langer}, {Goldsmith}, {Kuiper}, {Fischer}, {Gong}, {Brunthaler}, {Rugel}, \&
  {Menten}}]{2024ApJ...973...89P}
{Pineda}, J.~L., {Horiuchi}, S., {Anderson}, L.~D., {et~al.} 2024, \apj, 973,
  89, \dodoi{10.3847/1538-4357/ad615a}

\bibitem[{{Poglitsch} {et~al.}(2010){Poglitsch}, {Waelkens}, {Geis},
  {Feuchtgruber}, {Vandenbussche}, {Rodriguez}, {Krause}, {Renotte}, {van
  Hoof}, {Saraceno}, {Cepa}, {Kerschbaum}, {Agn{\`e}se}, {Ali}, {Altieri},
  {Andreani}, {Augueres}, {Balog}, {Barl}, {Bauer}, {Belbachir}, {Benedettini},
  {Billot}, {Boulade}, {Bischof}, {Blommaert}, {Callut}, {Cara}, {Cerulli},
  {Cesarsky}, {Contursi}, {Creten}, {De Meester}, {Doublier}, {Doumayrou},
  {Duband}, {Exter}, {Genzel}, {Gillis}, {Gr{\"o}zinger}, {Henning},
  {Herreros}, {Huygen}, {Inguscio}, {Jakob}, {Jamar}, {Jean}, {de Jong},
  {Katterloher}, {Kiss}, {Klaas}, {Lemke}, {Lutz}, {Madden}, {Marquet},
  {Martignac}, {Mazy}, {Merken}, {Montfort}, {Morbidelli}, {M{\"u}ller},
  {Nielbock}, {Okumura}, {Orfei}, {Ottensamer}, {Pezzuto}, {Popesso},
  {Putzeys}, {Regibo}, {Reveret}, {Royer}, {Sauvage}, {Schreiber}, {Stegmaier},
  {Schmitt}, {Schubert}, {Sturm}, {Thiel}, {Tofani}, {Vavrek}, {Wetzstein},
  {Wieprecht}, \& {Wiezorrek}}]{2010A&A...518L...2P}
{Poglitsch}, A., {Waelkens}, C., {Geis}, N., {et~al.} 2010, \aap, 518, L2,
  \dodoi{10.1051/0004-6361/201014535}

\bibitem[{{Pontoppidan} {et~al.}(2014){Pontoppidan}, {Salyk}, {Bergin},
  {Brittain}, {Marty}, {Mousis}, \& {{\"O}berg}}]{2014prpl.conf..363P}
{Pontoppidan}, K.~M., {Salyk}, C., {Bergin}, E.~A., {et~al.} 2014, in
  Protostars and Planets VI, ed. H.~{Beuther}, R.~S. {Klessen}, C.~P.
  {Dullemond}, \& T.~{Henning}, 363

\bibitem[{{Russeil} {et~al.}(2017){Russeil}, {Zavagno}, {M{\`e}ge}, {Poulin},
  {Molinari}, \& {Cambresy}}]{2017A&A...601L...5R}
{Russeil}, D., {Zavagno}, A., {M{\`e}ge}, P., {et~al.} 2017, \aap, 601, L5,
  \dodoi{10.1051/0004-6361/201730540}

\bibitem[{{Schinnerer} \& {Leroy}(2024)}]{2024ARA&A..62..369S}
{Schinnerer}, E., \& {Leroy}, A.~K. 2024, \araa, 62, 369,
  \dodoi{10.1146/annurev-astro-071221-052651}

\bibitem[{{Schmidt}(1959)}]{1959ApJ...129..243S}
{Schmidt}, M. 1959, \apj, 129, 243, \dodoi{10.1086/146614}

\bibitem[{{Shirley}(2015)}]{2015PASP..127..299S}
{Shirley}, Y.~L. 2015, \pasp, 127, 299, \dodoi{10.1086/680342}

\bibitem[{{Soler} {et~al.}(2023){Soler}, {Zari}, {Elia}, {Molinari}, {Mininni},
  {Schisano}, {Traficante}, {Klessen}, {Glover}, {Hennebelle}, {Colman},
  {Frankel}, \& {Wenger}}]{2023A&A...678A..95S}
{Soler}, J.~D., {Zari}, E., {Elia}, D., {et~al.} 2023, \aap, 678, A95,
  \dodoi{10.1051/0004-6361/202347608}

\bibitem[{{Solomon} {et~al.}(1987){Solomon}, {Rivolo}, {Barrett}, \&
  {Yahil}}]{1987ApJ...319..730S}
{Solomon}, P.~M., {Rivolo}, A.~R., {Barrett}, J., \& {Yahil}, A. 1987, \apj,
  319, 730, \dodoi{10.1086/165493}

\bibitem[{{Tacconi} {et~al.}(2020){Tacconi}, {Genzel}, \&
  {Sternberg}}]{2020ARA&A..58..157T}
{Tacconi}, L.~J., {Genzel}, R., \& {Sternberg}, A. 2020, \araa, 58, 157,
  \dodoi{10.1146/annurev-astro-082812-141034}

\bibitem[{{Urquhart} {et~al.}(2024){Urquhart}, {K{\"o}nig}, {Colombo},
  {Karska}, {Wyrowski}, {Menten}, {Moore}, {Brand}, {Elia}, {Giannetti},
  {Leurini}, {Figueira}, {Lee}, \& {Dumke}}]{2024MNRAS.528.4746U}
{Urquhart}, J.~S., {K{\"o}nig}, C., {Colombo}, D., {et~al.} 2024, \mnras, 528,
  4746, \dodoi{10.1093/mnras/stad3983}

\bibitem[{{Veneziani} {et~al.}(2013){Veneziani}, {Elia}, {Noriega-Crespo},
  {Paladini}, {Carey}, {Faimali}, {Molinari}, {Pestalozzi}, {Piacentini},
  {Schisano}, \& {Tibbs}}]{2013A&A...549A.130V}
{Veneziani}, M., {Elia}, D., {Noriega-Crespo}, A., {et~al.} 2013, \aap, 549,
  A130, \dodoi{10.1051/0004-6361/201219570}

\bibitem[{{Veneziani} {et~al.}(2017){Veneziani}, {Schisano}, {Elia},
  {Noriega-Crespo}, {Carey}, {Di Giorgio}, {Fukui}, {Maiolo}, {Maruccia},
  {Mizuno}, {Mizuno}, {Molinari}, {Mottram}, {Moore}, {Onishi}, {Paladini},
  {Paradis}, {Pestalozzi}, {Pezzuto}, {Piacentini}, {Plume}, {Russeil}, \&
  {Strafella}}]{2017A&A...599A...7V}
{Veneziani}, M., {Schisano}, E., {Elia}, D., {et~al.} 2017, \aap, 599, A7,
  \dodoi{10.1051/0004-6361/201423474}

\bibitem[{{Vutisalchavakul} \& {Evans}(2013)}]{2013ApJ...765..129V}
{Vutisalchavakul}, N., \& {Evans}, Neal~J., I. 2013, \apj, 765, 129,
  \dodoi{10.1088/0004-637X/765/2/129}

\bibitem[{{Whittet} {et~al.}(2013){Whittet}, {Poteet}, {Chiar}, {Pagani},
  {Bajaj}, {Horne}, {Shenoy}, \& {Adamson}}]{2013ApJ...774..102W}
{Whittet}, D.~C.~B., {Poteet}, C.~A., {Chiar}, J.~E., {et~al.} 2013, \apj, 774,
  102, \dodoi{10.1088/0004-637X/774/2/102}

\bibitem[{{Yusef-Zadeh} {et~al.}(2009){Yusef-Zadeh}, {Hewitt}, {Arendt},
  {Whitney}, {Rieke}, {Wardle}, {Hinz}, {Stolovy}, {Lang}, {Burton}, \&
  {Ramirez}}]{2009ApJ...702..178Y}
{Yusef-Zadeh}, F., {Hewitt}, J.~W., {Arendt}, R.~G., {et~al.} 2009, \apj, 702,
  178, \dodoi{10.1088/0004-637X/702/1/178}

\bibitem[{{Zari} {et~al.}(2023){Zari}, {Frankel}, \&
  {Rix}}]{2023A&A...669A..10Z}
{Zari}, E., {Frankel}, N., \& {Rix}, H.-W. 2023, \aap, 669, A10,
  \dodoi{10.1051/0004-6361/202244194}

\end{thebibliography}


\end{document}